\documentstyle[12pt,epsfig,amsmath]{article}
\newcommand {\be}{\begin{equation}}
\newcommand {\ee}{\end{equation}}
\newcommand {\ba}{\begin{eqnarray}}
\newcommand {\ea}{\end{eqnarray}}
\begin{document}

\def \a'{\alpha'}
\baselineskip 0.65 cm
\begin{flushright}
IPM/P-2007/007\\
\today
\end{flushright}

\begin{center}{\large
{\bf Electron Electric Dipole Moment from Lepton Flavor
Violation}} {\vskip 0.5 cm} {\bf Seyed Yaser Ayazi and Yasaman
Farzan }{\vskip 0.5 cm }
Institute for studies in Theoretical Physics and Mathematics (IPM)\\
P.O. Box 19395-5531, Tehran, Iran\\
\end{center}

\begin{abstract}
The general Minimal Supersymmetric Standard Model introduces new
sources for   Lepton Flavor Violation (LFV) as well as
CP-violation. In this paper, we show that when both sources are
present, the electric dipole moment  of the electron, $d_e$,
receives a contribution from the phase of the trilinear $A$-term
of staus, $\phi_{A_\tau}$. For $\phi_{A_\tau}=\pi/2$, the value of
$d_e$, depending on the ratios of the LFV mass elements,  can
range between zero and three orders of magnitude above the present
bound. We show that the present bound on $d_e$ rules out a large
portion of the CP-violating and the LFV  parameter space which is
consistent with the bounds on the LFV rare decays.
 We  show that   studying the correlation between $d_e$ and the  P-odd
 asymmetry in $\tau \to e\gamma$
helps us to derive a more conclusive bound on $\phi_{A_\tau}$.
  We also discuss the possibility of cancelation among the
contributions of different CP-violating phases to $d_e$.
\end{abstract}

\section{Introduction}
As is well-known, elementary particles can possess electric dipole
moments  only if CP is violated. In the framework of the Standard
Model (SM) of the elementary particles, the 3$\times$3 quark
mixing matrix (the CKM matrix) can accommodate a CP-violating
phase. In fact, CP-symmetry in the meson system has been observed
to be violated in accord with the SM. However, the effect of the
CP-violating phase of the CKM matrix on the Electric Dipole Moment
(EDM)
 of the
electron, $d_e$, is shown to be very small \cite{deneutrino} and
beyond the reach of  experiments  in the foreseeable future
\cite{ongoing,forthcoming}. Thus if the forthcoming
\cite{forthcoming} or   proposed experiments detect a nonzero
$d_e$, it will be an indication of physics beyond the SM.

Recent neutrino data proves that Lepton Flavor (LF) has been
violated in nature. The effect can be explained by mixing in the
neutrino mass matrix. In principle such a LF Violation (LFV) in
neutrino mass matrix can lead to  the LFV decays, $\tau \to
e\gamma$, $\tau \to \mu \gamma$ and $\mu \to e \gamma$
\cite{LFVneutrino}. However, if the neutrino mass matrix is the
only source of LFV, the branching ratio of these decays will be so
small that cannot be detected in foreseeable future. In the
future, if experiments  report a nonzero branching ratio for any
of the aforementioned LFV decays \cite{bfactory}, we will conclude
that the SM has to be augmented to include more sources of LFV. In
the context of Minimal Supersymmetric Standard Model (MSSM), which
is arguably the most popular extension of the SM, there are
several sources for CP-violation as well as for LFV which can lead
to effects exceeding the present experimental bounds. The bounds
on Br$(\ell_j \to \ell_i \gamma)$ constrain the sources of LFV in
the MSSM. Moreover, the bounds on the EDM of the elementary
particles constrain the CP-violating phases of the MSSM. For
vanishing  LFV sources, the bounds from the EDMs on the
CP-violating phases of MSSM parameters have been extensively
studied in the literature (for an incomplete list see)
\cite{cancelation,Reconciling,without cancelation}. In
\cite{Bartl}, the effects of the phases of LFV masses as well as
the LFV trilinear $A$-couplings on $d_e$ have been studied.
However, \cite{Bartl} does not discuss the possible effects of the
phase of $A_\tau$ (the trilinear supersymmetry breaking coupling
of stau). Notice that although $A_\tau$ is a LF conserving
coupling which deals only with the staus, in the presence of LFV,
it can affect the properties of leptons of other generations.

 In this paper, taking into account the possibility of the
LFV in soft supersymmetry breaking terms, we will focus on the
possible effects of the phase of $A_\tau$  on the electric dipole
moment of the electron. As is well-known, the phase of $A_\tau$,
$\phi_{A_\tau}$, can also manifest itself in the decay and
production of staus \cite{decay}. One of the goals of the proposed
state-of-the-art ILC project is detecting such effects \cite{ILC}.
It is therefore very exciting to learn about the value of
$\phi_{A_\tau}$ by present or forthcoming low energy experiments.

We show that for ${\rm Br}(\tau \to e \gamma)$ close to its
present bound, the bound on $d_e$ can already  constrain
$\phi_{A_\tau}$. We discuss how other sets of the CP-violating
phases can mimic the effect of $\phi_{A_\tau}$ on $d_e$ and
suggest some ways to resolve the degeneracies. Recently, it has
been shown in \cite{tracing} that by measuring the spin of the
final particles in the LFV rare decays, one can extract
information on the CP-violating phases of the underlying theory.
In the present paper, we however do not take into account such a
possibility and focus on the spin-averaged decays rate.

The paper is organized as follows. In Sec. 2, we specify the model
that we study in this paper and summarize the observable effects
that can be used to extract information on the parameters of the
model. In Sec. 3, we discuss how CP-violating and LFV parameters
affect $d_e$ and  other observable quantities and present scatter
plots that explore the parameter space. In Sec. 4, we first
enumerate the possible CP-violating phases and evaluate their
respective effects
 with special emphasis on the possibility of
cancelation. Section 5, summarizes our conclusions. The formulae
for calculating the rate of LFV rare decays and $d_e$ are
summarized in the appendix.
\section{The model and its observable effects}
In this section we specify the model and the sources of LFV and
CP-violation that we are going to study.

 In this paper, we
consider the Minimal Supersymmetric Standard Model with
superpotential
 \be \label{superpotential}
   W_{{\rm MSSM}} =
           - Y_{i}  \widehat{{e}^c_{Ri}} \ \widehat{L}_i  \cdot \widehat{H_{d}}
-\mu\ \widehat{H_{u}}\cdot \widehat{H_{d}} \ee
 where $\widehat{L}_i$, $\widehat{H_{u}}$ and $ \widehat{H_{d}}$
 are doublets of chiral superfields respectively associated  with
 left-handed leptons and the two Higgs doublets of the
 MSSM. In the above formula,
  $\widehat{{e}^c_{Ri}}$ is
the chiral superfields associated with the  right-handed charged
lepton fields. The index $``i"$ determines the flavor; $ i=1,2,3$
respectively correspond to $ e,\mu,\tau $. Notice that we have
chosen the mass basis for the charged leptons ({\it i.e.,} Yukawa
coupling of the charged leptons is taken to be diagonal). Notice
that the Yukawa terms involving the quark supermultiplets have to
be added to (\ref{superpotential}). However, since we are going to
concentrate on the lepton sector, we have omitted such terms. At
the electroweak scale, the soft supersymmetry breaking part of the
Lagrangian in general has the form \ba \label{MSSMsoft}\L_{\rm
soft}^{\rm MSSM} &=&-\ 1/2 \left(  M_1 \widetilde{B}\widetilde{B}+
M_2 \widetilde{W} \widetilde{W} +{\rm H.c.} \right) \cr
&-&(A_{i}Y_{ i}\delta_{ij}+A_{ij}) \widetilde{e_{Ri}^c} \
\widetilde{L_{j}} \cdot H_{d} + {\rm H.c.} ) - \widetilde{L_{i}}
^{\dag} \ ( m_{\tilde{e}_{L } }^{2})_{ij}\widetilde{L_{j}} -
\widetilde{e_{Ri}^c }^{\dag} \ (m_{\tilde{e}_{R
}}^{2})_{ij}\widetilde{e_{Rj}^c}\cr &-& \ m_{H_{u}}^{2}\
H_{u}^{\dag}\ H_{u}-\ m_{H_{d}}^{2}\ H_{d}^{\dag}\ H_{d}-(\ B_H \
H_{u}\cdot H_{d}+ {\rm H.c.}),\ea where the ``$i$" and ``$j$"
indices determine the flavor and $\tilde{L}_i$ consists of
$(\tilde{\nu}_i \ \tilde{e}_{Li})$. Notice that we have divided
the trilinear coupling to a flavor diagonal  part
($A_{i}Y_{i}\delta_{ij}$) and a LFV part ($A_{ij}$ with
$A_{ii}=0$). Again terms involving the squarks as well as the
gluino mass term have to be added to (\ref{MSSMsoft}). The
Hermiticity of the Lagrangian implies that $m_{H_{u}}^2$,
$m_{H_{d}}^2$ and the diagonal elements of $m_{\tilde{e}_L}^2$ and
$m_{\tilde{e}_{R }}^2$ are all real. Moreover, without loss of
generality we can rephase the fields to make the parameters $M_2$,
$B_H$ as well as $Y_{ i}$ real. In such a basis, the rest of above
parameters can in general be complex and can be considered as
sources of CP-violation  giving contributions to EDMs.

After electroweak symmetry breaking, the $A$-terms in
(\ref{MSSMsoft}) as well as the terms in  superpotential  induce
left-right mixing. The Hermitian 6$\times$6 mass matrix of
$(\tilde{e}_R)_i$ and $(\tilde{e}_L)_j$ can  in general be
written in terms of three 3$\times 3$ mass sub-matrices $m_L^2$,
$m_R^2$ and $m_{LR}^2$ as follows
\begin{equation}
L_{{\rm slepton}}=
-\left(%
\begin{array}{cc}
  \widetilde{e}_L^\dagger& \widetilde{e}_R^\dagger \\
\end{array}%
\right)\left(%
\begin{array}{cc}
  m_L^2 & m_{LR}^{2\dag} \\
  m_{LR}^2 & m_R^2 \\
\end{array}\right)\left(%
\begin{array}{c}
  \widetilde{e}_L \\
  \widetilde{e}_R \\
\end{array}%
\right).
 \end{equation}
The formulae for  $m_L^2$, $m_R^2$ and $m_{LR}^2$ in terms of the
soft supersymmetry breaking potential  are given in
Eqs.~(\ref{mL},\ref{mR},\ref{mLR}) of the appendix.
 With above Lagrangian and superpotential, the LF is conserved if
and only if $A_{ij}=0$ and the off-diagonal elements of
$m_{\tilde{e}_L}^2 $ and $m_{\tilde{e}_R}^2$ vanish (for $i\ne j$,
$(m_L^2)_{ij}=(m_R^2)_{ ij}=A_{ij}=0$). At the one loop level, in
the lepton conserving case, each $A$ term can contribute to the
EDM of only the corresponding fermion. For example, at the one
loop level, the phase of $A_\tau$ will have no effect on $d_e$ but
can induce an EDM for the tau lepton of order of ${\rm
Im}(A_\tau)m_\tau/m_{{\rm SUSY}}^3$. Considering
 the fact that the bound on the EDM of tau is much weaker than
 this
\cite{pdg}, no bound on $\phi_{A_\tau}$ from $d_\tau$ can be
derived. In the LF conserving case, $d_e$ will receive significant
contributions from the phases of $A_e$, $\mu$ and $M_1$. Thus, the
strong bound on $d_e$ can be translated into bounds on the phases
of these parameters. In the literature, it is shown that for
relatively low scale supersymmetry ($m_{{\rm SUSY}}< 500$~GeV),
the bounds on these phases from $d_e$ are very strong
\cite{without cancelation} unless severe cancelation takes place
\cite{cancelation}.

At the two-loop level, even in the lepton flavor conserving case,
the phase of $A_{\tau}$ can induce a contribution to $d_e$ as well
as to $d_n$ \cite{Chang:1998uc}\footnote{Although in
\cite{Chang:1998uc} the two-loop effects of only $\rm {Im[A_b]}$
and $\rm{Im[A_t]}$ on $d_e$ and $d_n$ have been discussed, similar
discussion also holds for $\rm{Im[A_{\tau}]}$.}. For relatively
large values of $\tan{\beta}$ ($\tan{\beta}\geq10$) and $m_{\rm
SUSY}\simeq100~\rm {GeV}$, the bound on $d_e$ can be translated
into a bound of order of few hundred $\rm{GeV}$ on
$\rm{Im[A_{\tau}]}$. The limit from the bound on $d_n$ even is
less stringent \footnote{ We would like to thank the anonymous
referee for pointing out such a possibility.}.

For LFV case, the $A$-term associated with a definite lepton
flavor can in principle affect the EDM of a lepton of another
flavor, even at the one-loop level. In particular if the $e\tau$
element of $m^2_{L}$ and $m^2_{R}$ or $A_{ij}$ are nonzero, the
phase of $A_\tau$ can induce an EDM for the electron exceeding the
present bound by several orders of magnitudes. As a result, in
this case the strong bound on $d_e$ can severely restrict the
phase of $A_\tau$. In order to study this bound, we have to first
consider the bounds on the LFV masses and $A$-terms from the
bounds on the LFV decay modes of the charged leptons. Notice that
throughout this paper we have implicitly assumed that the origin
of LFV lies at an energy scale far above the scale of the
electroweak symmetry breaking. We therefore have the same
LFV-violating elements for the left-handed charged lepton and
sneutrino mass matrices.

The strongest upper bound on the LFV elements of the slepton mass
matrices comes from the following experimental bound: \be
\label{br(mu->e gamma)} {\rm Br}(\mu\to e \gamma)<1.2\times
10^{-11} \ee which for $m_{{\rm SUSY}}\sim 100~ {\rm GeV}$ implies
$(m_{L}^2)_{e \mu},(m_{R}^2)_{e \mu}
\stackrel{<}{\sim}10^{-4}-10^{-3}(m_{{\rm SUSY}}^2) $ and $A_{e
\mu},A_{\mu e}\stackrel{<}{\sim} 0.05 m_{{\rm SUSY}} ^2/\langle
H_d \rangle.$ Throughout this paper we will set these LFV elements
equal to zero:
$$(m_{L}^2)_{e \mu}=(m_{R}^2)_{e \mu}=0  \ \ {\rm and}
\ \ A_{e\mu}=A_{\mu e}=0.$$ There are also strong bounds on the
branching ratios of LFV decay modes of the tau lepton: \be
\label{br(tau->e gamma)} {\rm Br}(\tau \to e \gamma)< 9.4 \times
10^{-8}\ \cite{banerjee}\ee and \be \label{br(tau->mu gamma)} {\rm
Br}(\tau \to \mu \gamma)< 1.6 \times 10^{-8}\ \cite{banerjee}\ee
which can be respectively translated into bounds on the $\tau e$-
and $\tau \mu$-elements. However, it can be shown that the bound
on the $\tau e$-elements from (\ref{br(tau->e gamma)}) are not
very strong and these elements can be of the same order as the
diagonal elements. Suppose that both the $\tau e$- and $\tau
\mu$-elements are nonzero. This means the $e$- and $\mu$-lepton
numbers are both violated and the $\mu \to e \gamma$ decay can
therefore take place despite the vanishing $\mu e$-elements. In
fact, for relatively large $\tau e$-elements saturating the bound
from (\ref{br(tau->e gamma)}), the bound (\ref{br(mu->e gamma)})
can be translated into a strong bound on the $\tau \mu $-elements
which is more stringent than the bound from ${\rm Br}(\tau \to \mu
\gamma)$. Throughout this paper, we will set all the $\tau \mu$
equal to zero:
$$(m_{L
}^2)_{\mu \tau}=(m_{R}^2)_{\mu \tau}=0  \ \ {\rm and} \ \
A_{\mu\tau}=A_{\tau \mu}=0.$$ In this scenario, the $\mu$-flavor
number is thus conserved.

As shown in the literature, integrating out the heavy
supersymmetric states,  $\tau \to e \gamma$ can be described by
the following effective Lagrangian \be \label{effectiveL} e
\epsilon_\alpha^\dagger m_\tau q_\beta
\left[\bar{e}_R\sigma^{\alpha \beta}(A_L)_{e\tau}
\tau_L+\bar{e}_L\sigma^{\alpha
\beta}(A_R)_{e\tau}\tau_R\right]+{\rm H.c.} \ee where
$\epsilon_\alpha$ is the photon field and $q_\beta$ is its
four-momentum and $\sigma^{\alpha
\beta}=\frac{i}{2}[\gamma^\alpha,\gamma^\beta]$. The formulae for
$A_L$ and $A_R$ in terms of the supersymmetric parameters have
been derived in \cite{hisano} for the CP-invariant case. We have
re-derived the formulae for the CP-violating case. The results can
be found in the appendix. Our results are in agreement with
\cite{hisano} in the CP-invariant limit. Using (\ref{effectiveL})
it is straightforward to show that in the rest frame of the tau
lepton, the partial decay rate is given by \be \label{partial}
\frac{d\Gamma(\tau \to e \gamma)}{d\cos \theta}={e^2 \over
32\pi}m_\tau^5\left[ (|(A_L)_{e\tau}|^2(1+\cos \theta)+|(A_R)_{e
\tau}|^2(1-\cos \theta)\right]
 \ee
 where $\theta$ is the angle between the  spin of the tau and the
 momentum of the emitted electron. Integrating over $\theta$ we
 obtain
 \be \label{total}
 \Gamma(\tau \to e \gamma)=\frac{e^2}{16 \pi}
 m_\tau^5(|(A_L)_{e\tau}|^2+|(A_R)_{e\tau}|^2). \ee
 Notice that different sets of LFV  mass matrix elements can
 result in the same rate for $\tau \to e \gamma$.  Let us define the $A_P$ asymmetry   as
 follows \be \label{apdef}
 A_P=4\times {\int_0^1 \frac{d \Gamma(\tau\to e \gamma)}{d\cos
 \theta}d\cos \theta-\int_{-1}^0 \frac{d \Gamma(\tau\to e \gamma)}{d\cos
 \theta}d\cos \theta \over \Gamma(\tau \to e \gamma)}. \ee
Using (\ref{partial}), we can prove that
$$A_P={|(A_L)_{e\tau}|^2-|(A_R)_{e\tau}|^2 \over |(A_L)_{e\tau}|^2+|(A_R)_{e\tau}|^2}.
$$
Thus  by measuring the partial decay rate of  $\tau$, we will be
able to extract not only $(|(A_L)_{e\tau}|^2+|(A_R)_{e\tau}|^2)$
but also $(|(A_L)_{e\tau}|^2-|(A_R)_{e\tau}|^2)$. In
\cite{kitano}, it has been shown that by studying the angular
distributions of the final particles at an $e^-e^+$ collider such
as a B-factory, it will be possible to derive $A_P$. $A_P$
provides us with more information on the LFV parameters of the
underlying theory. For example, if the source of LFV is a
canonical seesaw mechanism embedded in the MSSM, we expect
$(m_R^2)_{e \tau}\ll (m_L^2)_{e \tau}$ and $A_{ e\tau}\ll A_{ \tau
e}$ which means $|(A_L)_{e\tau}|^2\ll |(A_R)_{e\tau}|^2$ and
therefore $A_P \to -1$. In this paper, we will study the
correlation between $d_e$, ${\rm Br}(\tau \to e \gamma)$ and $A_P$
and discuss the possibility of resolving the degeneracies by
combining  the information on their values.

\section{New contributions to $d_e$ in the presence of LFV}

\begin{figure}[h]\begin{center}
  \includegraphics[height= 5 cm,bb=45 125 380 350,clip=true]{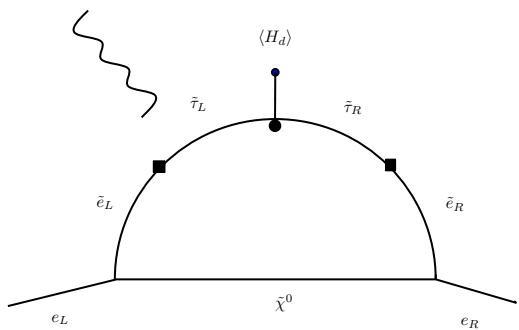}
  \caption{A neutralino exchange diagram contributing to $d_e$. The photon can attach to any of
  the $\tilde{e}_L$, $\tilde{\tau}_L$, $\tilde{\tau}_R$ or $\tilde{e}_R$ propagators. The boxes
  on the left and right sides respectively depict insertion of $(m_L^2)_{e\tau}$ and
  $(m_R^2)_{e\tau}$. The circles indicate insertion of the $A_\tau$ vertex and the vacuum expectation
  value of $H_d$.  } \label{massinsertion}
\end{center}\end{figure}

Let us for the moment suppose $A_{e\tau}=A_{\tau e}=0$. As
illustrated in Fig.~\ref{massinsertion}, for nonzero $(m_R^2)_{
e\tau}$ and $(m_L^2)_{\tau e}$, the phase of $A_\tau$ can induce a
contribution to $d_e$. As a result for definite values of the
off-diagonal mass elements, the bound on $d_e$ can be interpreted
as a bound on $\phi_{A_\tau}$ or on ${\rm Im}(A_\tau)$. Consider
the case that both ($m^2_R)_{\tau e}$ and $(m_L^2)_{\tau e}$ are
close to the corresponding bounds from Br$(\tau \to e
 \gamma) $.  In this case, ${\rm Br}(\tau \to e \gamma)$ is close to
its present bound and $A_P$  takes a value in the interval (-1,1);
{\it i.e.,} $A_P\ne \pm 1$. For such a configuration,
 we expect the bound on Im$(A_\tau)$ to
be more stringent than the bound on Im$(A_e)$ because the effect
of Im($A_\tau$) is given by $m_\tau {\rm Im}(A_\tau)(m^2_R)_{e\tau
}(m^2_L)_{e\tau}/m^7_{susy}$, whereas the effect of Im($A_e$) is
proportional to $m_e{ \rm Im}(A_e)/m^3_{susy}$. Now, suppose only
one of $(m_R^2)_{ e\tau}$ and $(m_L^2)_{\tau e}$ is close to its
upper bound from ${\rm Br}(\tau \to e \gamma)$ and the other is
zero or very small. In this case, $A_P$ will converge either to 1
(for $(m_R^2)_{e \tau}\gg (m_L^2)_{e \tau}$) or to -1 (for
$(m_L^2)_{e \tau} \gg (m_R^2)_{e \tau}$). From Fig.
\ref{massinsertion} we observe that if only one of $(m_R^2)_{
e\tau}$ or $(m_L^2)_{\tau e}$ is nonzero and the rest of the
$e\tau$ entrees (including $A_{e\tau}$ and $A_{\tau e}$) vanish,
at one-loop level, the phase of $A_\tau$ cannot contribute to
$d_e$.

\begin{figure}
\psfig{figure=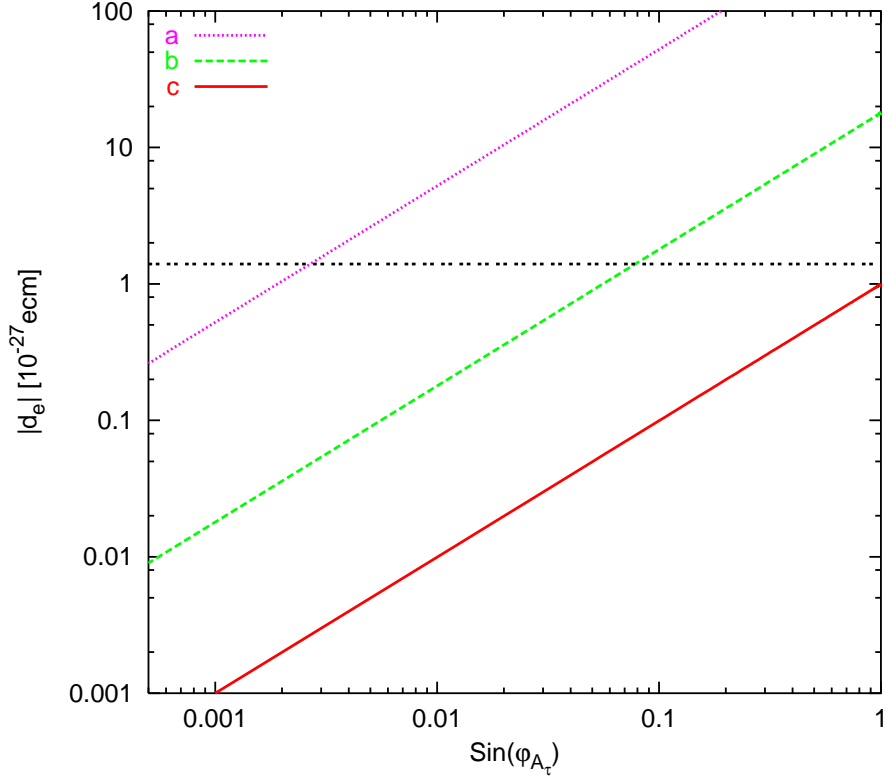,bb= 50 25 553 495, clip=true, height=4.5
in} \caption{$d_e$ versus $\sin \phi_{A_\tau}$. The input
parameters correspond to the $\alpha$ benchmark proposed in
\cite{NUHMbenchmark}: $|\mu|=375$~GeV, $m_0=210$~GeV, ${\rm
M}_{1/2}=285$~GeV and $\tan \beta=10$ and we have set
$|A_{\tau}|$=500~GeV. All the LFV elements of the slepton mass
matrix are set to zero except $(m_L^2)_{e \tau}$ and $(m_R^2)_{e
\tau}$. The dotted (pink) line labeled (a) corresponds to
$(m_L^2)_{e \tau}$=$3500~{\rm GeV}^2$ and $(m_R^2)_{e
\tau}$=$15000~{\rm GeV}^2$. The dashed (green) line labeled (b)
corresponds to $(m_L^2)_{e \tau}$=$50~{\rm GeV}^2$ and $(m_R^2)_{e
\tau}$=$37000~{\rm GeV}^2$. The solid (red) line labeled (c)
corresponds to $(m_L^2)_{e \tau}$=$3500~{\rm GeV}^2$ and
$(m_R^2)_{e \tau}$=$30~{\rm GeV}^2$. The horizontal doted line at
$1.4 \times 10^{-27}~e~{\rm cm}$ depicts the present experimental
 limit \cite{pdg} on $d_e$.} \label{forgotten}
\end{figure}

\begin{figure}
\begin{center}
\centerline{\epsfig{figure=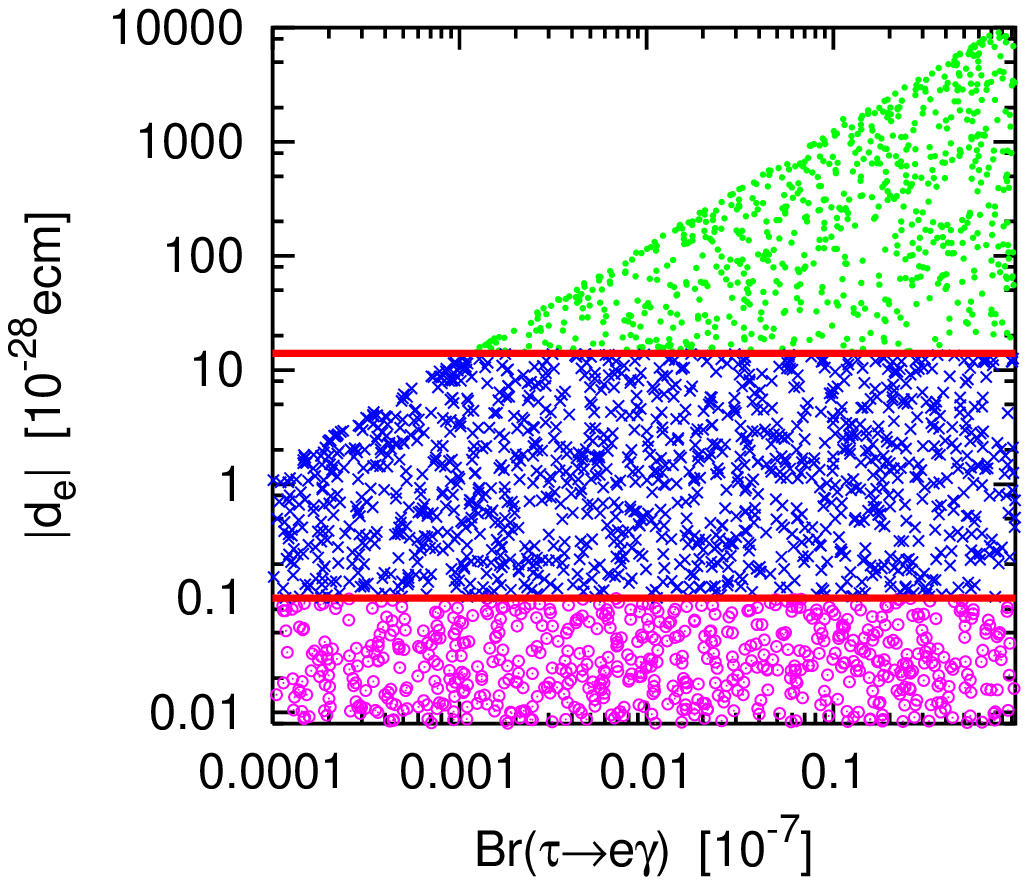,width=7.5cm}\hspace{5mm}\epsfig{figure=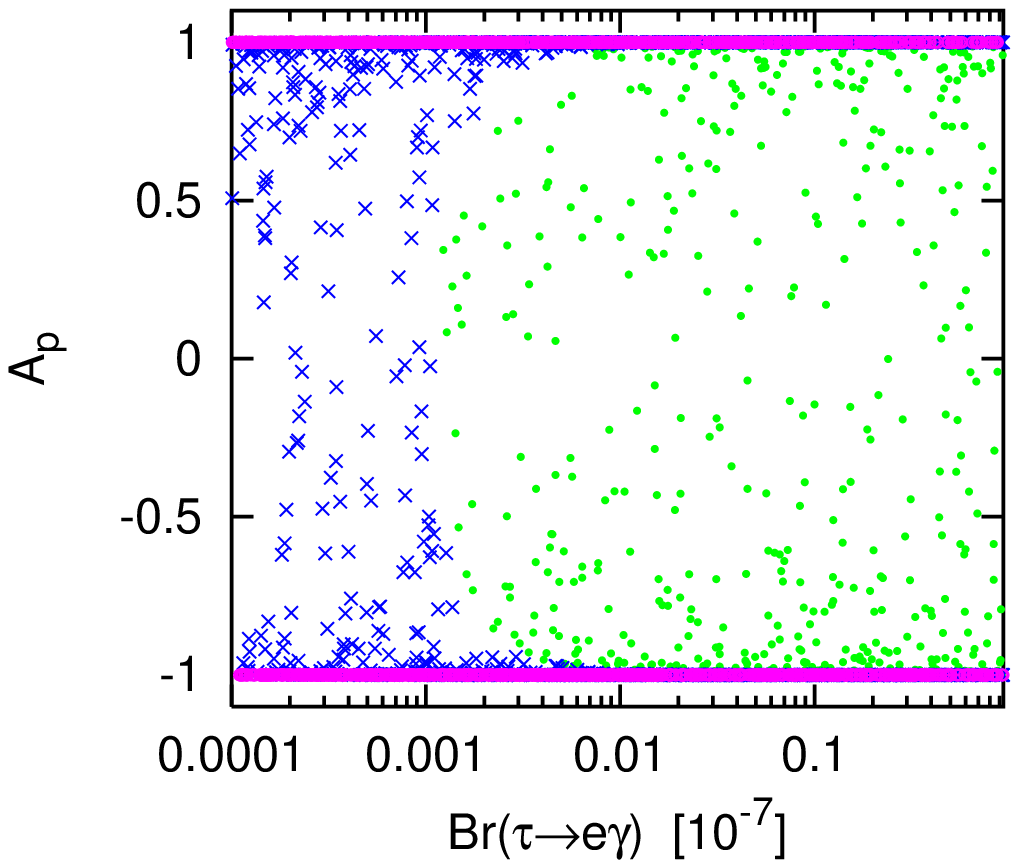,width=7.5cm}}
\centerline{\vspace{1.cm}\hspace{0.5cm}(a)\hspace{7cm}(b)}
\end{center}
\caption{ a) Scatter plot of $d_e$ versus ${\rm
Br}(\tau\rightarrow e\gamma)$. The input parameters correspond  to
the $\alpha$ benchmark proposed in \cite{NUHMbenchmark}:
$|\mu|=375$~GeV, $m_0=210$~GeV, ${\rm M}_{1/2}=285$~GeV and $\tan
\beta=10$. We have however set $\phi_{A_{\tau}}=\pi/2$ and
$|A_{\tau}|=500$~GeV. All the LFV elements of the slepton mass
matrix are set to zero except $(m^2_L)_{e\tau}$ and
$(m^2_R)_{e\tau}$ which pick up random values at a
 logarithmic scale respectively
from $(5.9\times 10^{-4}~{\rm GeV}^2,5.9 \times 10^{3}~{\rm
GeV}^2)$ and ($3.7 \times 10^{-3}~{\rm GeV}^2, 3.7\times
10^{4}~{\rm GeV}^2$).
 The horizontal line at $1.4 \times 10^{-27}~e~{\rm cm}$ depicts the present experimental
 limit \cite{pdg} and the one at $10^{-29}~e~{\rm cm}$ shows the limit that can be probed in the near future
 \cite{ongoing}.
b) Scatter plot of $A_P$ versus ${\rm Br}(\tau\rightarrow
e\gamma)$. For each scatter point in Fig.~\ref{alphaMM}-a there is
a counterpart in Fig.~\ref{alphaMM}-b corresponding to the same
input values for the $e\tau$ elements which is shown with the same
color and symbol. Notice that points shown in pink (corresponding
to $d_e<10^{-29}~e~{\rm cm}$) all lie on the $A_p$=$\pm1$.}
\label{alphaMM}
\end{figure}

\begin{figure}
\begin{center}
\centerline{\epsfig{figure=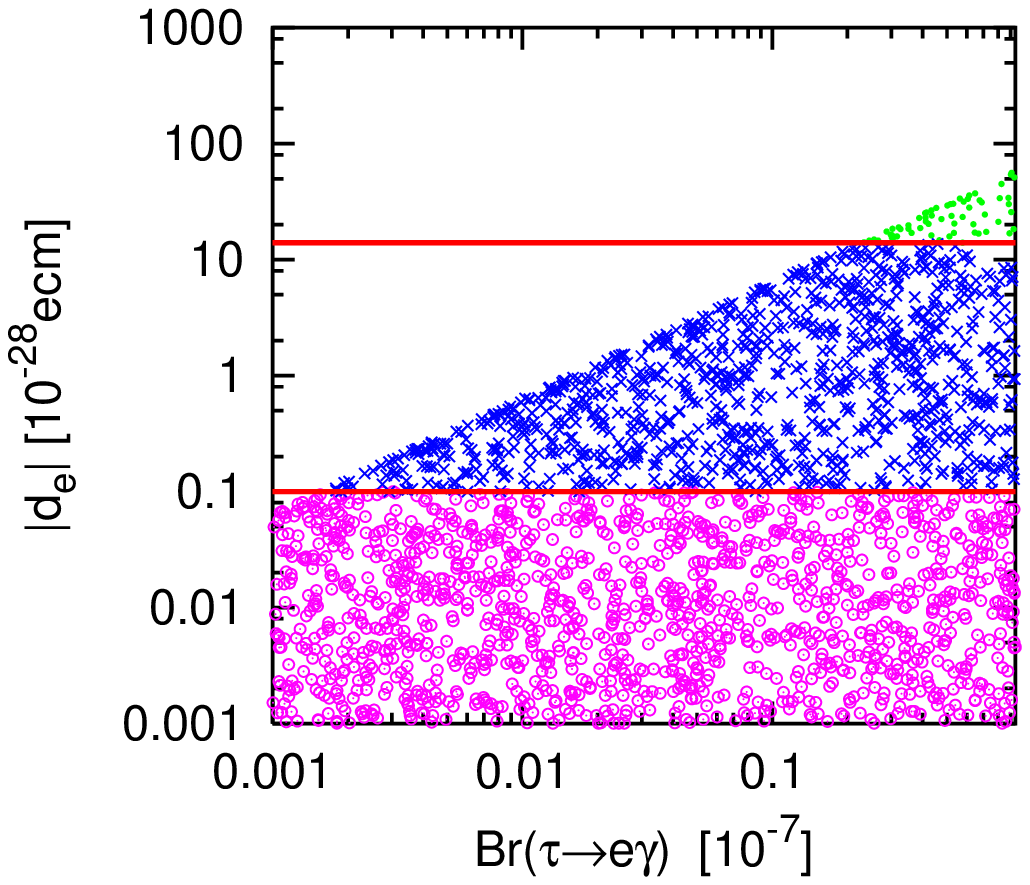,width=7.5cm}\hspace{5mm}\epsfig{figure=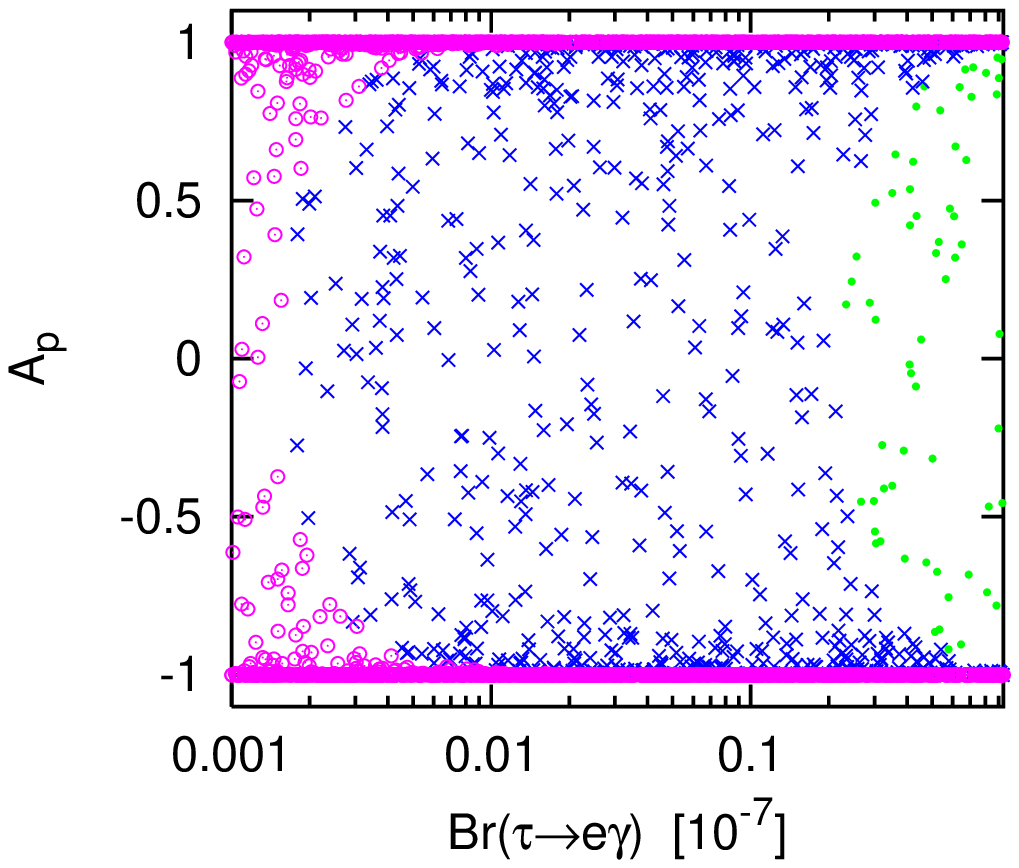,width=7.5cm}}
\centerline{\vspace{1.cm}\hspace{0.5cm}(a)\hspace{7cm}(b)}
\end{center}
\caption{Similar to Fig.~\ref{alphaMM} except that
$(m^2_L)_{e\tau}=(m^2_R)_{e\tau}=0$ and instead
$(m^2_{LR})_{e\tau}(=A_{e\tau}\langle H_d\rangle)$ and
$(m^2_{LR})_{\tau e}(=A_{ \tau e}\langle H_d\rangle)$ pick up
random values at a logarithmic scale from $(1.2\times 10^{-3}~{\rm
GeV}^2,1.2\times 10^{3}~{\rm GeV}^2)$. For each scatter point in
Fig.~\ref{alphaAA}-a there is a counterpart in
Fig.~\ref{alphaAA}-b corresponding to the same input values for
the $e\tau$ elements which is shown with the same color and
symbol.} \label{alphaAA}
\end{figure}

Figs.~\ref{forgotten}-\ref{alphaAA} demonstrate this observation.
To draw the figures in this paper, we have chosen the mass
spectrum corresponding to the $\alpha$  benchmark proposed in
\cite{NUHMbenchmark}. However, we have allowed the mass spectrum
of the staus to slightly deviate from these benchmarks. Notice
that at these benchmarks, the lightest stau is considerably
heavier than the lightest neutralino so stau-neutralino
coannihilation cannot play any significant role in fixing the dark
matter relic density. As a result, a slight change of stau
parameters will not dramatically affect the cosmological
predictions. Although for illustrative purposes we have displayed
the mass insertion approximation in Fig.~\ref{massinsertion}, to
calculate $d_e$ and ${\rm Br}(\tau \to e \gamma)$ we have used the
exact formulae (without the mass insertion approximation)
presented in the appendix.

Fig.~\ref{forgotten} depicts $d_e$ in terms of the sine of
$\phi_{A_\tau}$ for $A_{ij}=0$ and various values of $(m_L^2)_{e
\tau}$ and $(m_R^2)_{e \tau}$. This diagram demonstrates that for
$(m_L^2)_{e \tau}$ and $(m_R^2)_{e \tau}$ close to their bounds
from ${\rm Br}(\tau \to e \gamma)$, a very strong bound on
$\phi_{A_\tau}$ can be derived. That is while if there is a
hierarchy between these elements, the bound will be much weaker.
Notice that for the input parameters chosen for this figure, ${\rm
Br}(\tau \to e \gamma)$ lies close to its present bound: $10^{-8}
<{\rm Br}(\tau \to e \gamma)<10^{-7}$.

Fig.~\ref{alphaMM} demonstrates the correlation between $A_P$ and
$d_e$. As explained in the caption, the input mass spectrum is
that
 of the $\alpha$ benchmark \cite{NUHMbenchmark} and
$|A_\tau|=500$~GeV. We have set $A_{e \tau}$ and $A_{\tau e}$
equal to zero and the maximal value for the CP-violating phase is
chosen: $\phi_{A_\tau}=\pi/2$. $(m_L^2)_{\tau e}$ and
$(m_R^2)_{\tau e}$ pick up random values  at a logarithmic scales.
Points for which ${\rm Br}(\tau \to e \gamma)$ exceeds its present
bound are eliminated. Fig.~(\ref{alphaMM}-a) shows us that if
$\phi_{A_\tau}=\pi/2$, $d_e$ for a significant portion of the
scatter points exceeds the present bound. Fig.~(\ref{alphaMM}-b)
shows $A_P$ versus ${\rm Br}(\tau \to e \gamma)$ for the same
scatter points. To illustrate the correlation between $A_p$ and
$d_e$, we have shown the corresponding scatter points in
Fig~\ref{alphaMM}-a and Fig~\ref{alphaMM}-b with the same color
and symbol. That is at points marked with green dots, $d_e$
exceeds its present bound and at the scatter points marked with
blue cross ``$\times$" $10^{-29}<d_e<1.4\times 10^{-27}~e~{\rm
cm}$. The scatter points depicted by  pink circle, which appear in
Fig.~(\ref{alphaMM}-b) as two pink horizontal lines at $A_P=\pm
1$, correspond to $d_e<10^{-29}~e~{\rm cm}$. From
Fig.~(\ref{alphaMM}-b) we conclude that for $A_{e \tau}=A_{\tau
e}=0$, the bound on $d_e$ can be satisfied if either ${\rm
Br}(\tau \to e \gamma)$ is very small (which means that all the
LFV masses are very small) or $A_P$ is close to $\pm 1$ (meaning
that there is a hierarchy between the LFV elements). In other
words within this scenario, if future searches find $5 \times
10^{-10}<{\rm Br}(\tau \to e \gamma)$ and $-0.9<A_P<0.9$, the
bound on $d_e$ should be interpreted either as a bound on
$\phi_{A_\tau}$ or as an indication for a cancelation between
different contributions from $\phi_{A_\tau}$ and other possible
CP-violating phases. In the next section, we shall elaborate on
the latter possibility in more detail. We emphasize that to draw
this conclusion we have taken $A_{e \tau}=A_{\tau e}=0$.
 Repeating the same analysis with different benchmarks we have
found that the pattern and the correlation among ${\rm Br}(\tau
\to e \gamma)$, $A_P$ and $d_e$ are not sensitive to the input
values of LF conserving parameters.

\begin{figure}
\begin{center}
\centerline{\epsfig{figure=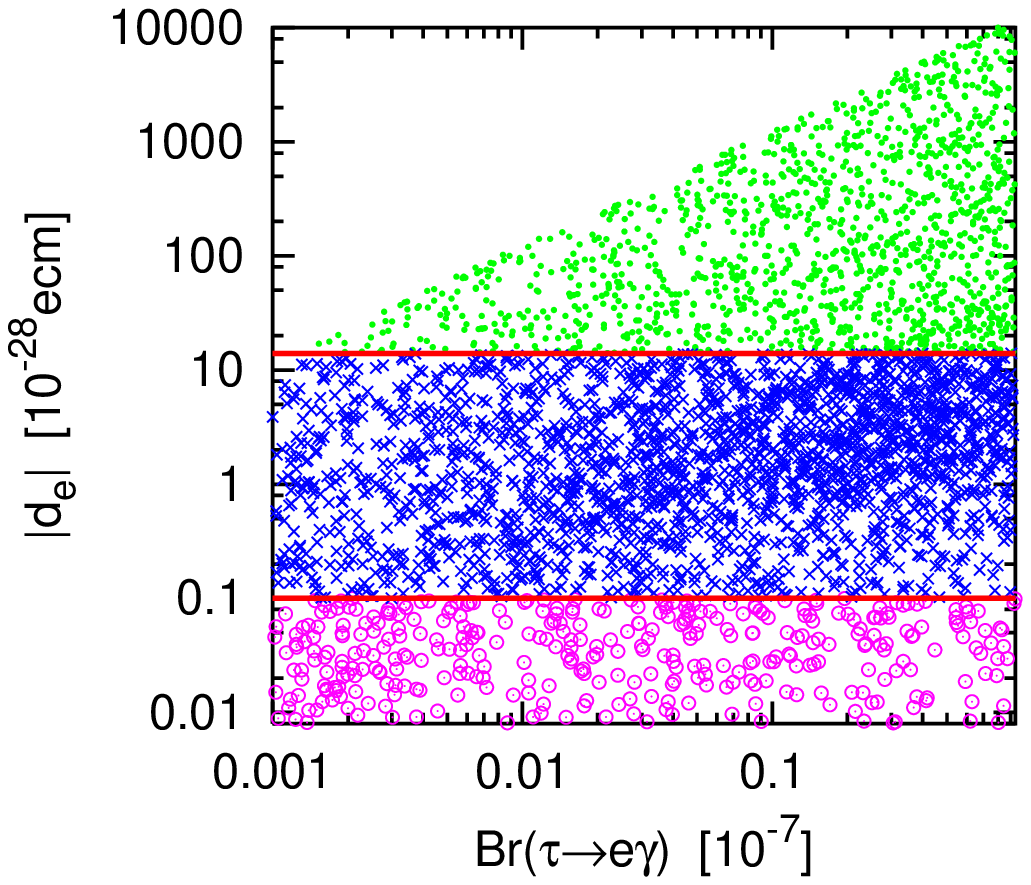,width=7.5cm}\hspace{5mm}\epsfig{figure=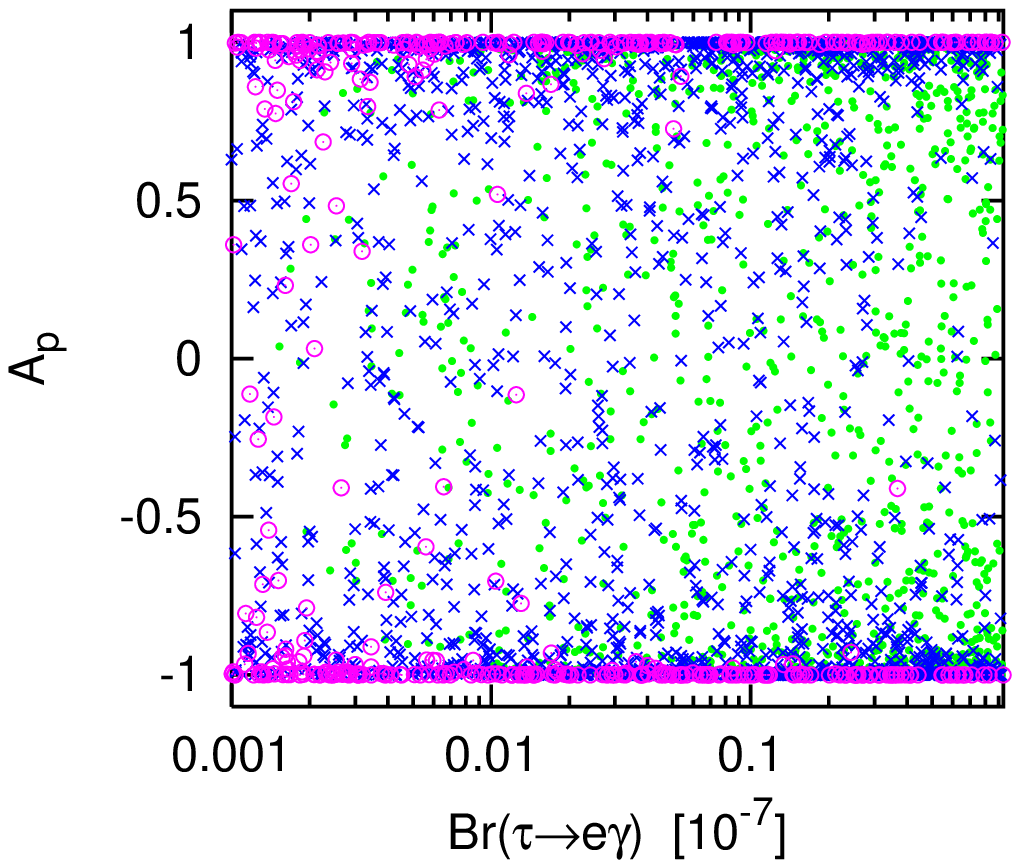,width=7.5cm}}
\centerline{\vspace{1.cm}\hspace{0.5cm}(a)\hspace{7cm}(b)}
\end{center}
\caption{Similar to Fig.~\ref{alphaMM} except that here in
addition to  $(m_L^2)_{e \tau}$ and $(m_R^2)_{e \tau}$,
$(m^2_{LR})_{e\tau}(=A_{e\tau}\langle H_d\rangle)$ and
$(m^2_{LR})_{\tau e}(=A_{ \tau e}\langle H_d\rangle)$ are allowed
to be nonzero. The values of $(m_L^2)_{e \tau}$ and $(m_R^2)_{e
\tau}$ are randomly chosen respectively from ($0.59~{\rm
GeV}^{2},5.9 \times 10^3~{\rm GeV}^2)$ and $(3.7~{\rm
GeV}^2,3.7\times 10^4~{\rm GeV}^2 )$ at a logarithmic scale.
$(m^2_{LR})_{e\tau }$ and $(m^2_{LR})_{\tau e}$ pick up random
values at a logarithmic scale from the interval $(0.12~{\rm
GeV}^2,1.2\times10^3~{\rm GeV}^2)$. For each scatter point in
Fig.~\ref{alphaMMAA}-a there is a counterpart in
Fig.~\ref{alphaMMAA}-b corresponding to the same input values for
the $e\tau$ elements which is shown with the same color and
symbol.} \label{alphaMMAA}
\end{figure}

To draw Fig.~\ref{alphaAA}, $(m_L^2)_{e \tau}$ and $(m_R^2)_{e
\tau}$ are set equal to zero and instead random values for
$A_{e\tau}$ and $A_{\tau e}$ are taken. The LF conserving
parameters for Figs.~(\ref{alphaMM}) and (\ref{alphaAA}) are the
same. Like the case of Fig. (\ref{alphaMM}), a significant portion
of the scatter points have $d_e$ exceeding the present bound.
Moreover a similar correlation between $d_e$, $A_P$ and ${\rm
Br}(\tau \to e \gamma)$ emerges. That is points marked with green
dot (corresponding to $d_e>$$1.4\times10^{-27}~e~{\rm cm})$, with
blue ``$\times$"  (corresponding to $10^{-29}<d_e<1.4\times
10^{-27}~e~{\rm cm}$) and pink circles (corresponding to
$d_e<10^{-29}~e~{\rm cm}$) are scattered respectively from right
to left. Notice however that in contrast to
Fig.~(\ref{alphaMM}-b), Fig.~(\ref{alphaAA}-b) includes scatter
points with $-0.9<A_P<0.9$ and ${\rm Br}(\tau \to e \gamma)\sim
10^{-8}$ that satisfy the present bound on $d_e$ (the points
marked with ``$\times$" in the plot). We have repeated the same
analysis with the $\delta$ benchmark and have found a similar
pattern. Similarity between the patterns
 means that the above
observation does not depend on the input values for the LF
conserving parameters.

In Fig.~\ref{alphaMMAA}, the input values for the LF conserving
parameters are taken to be the same as those for
Figs.~\ref{alphaMM} and \ref{alphaAA}. However, $(m_L^2)_{e\tau}$,
$(m_R^2)_{e \tau}$, $A_{e\tau}$ and $A_{\tau e}$ all take nonzero
random values. Fig.~(\ref{alphaMMAA}-a) contains features of both
Figs.~(\ref{alphaMM}) and (\ref{alphaAA}). The significant point
is that setting all the $e \tau$ mass elements nonzero, the
correlation among $A_P$, $d_e$ and ${\rm Br}(\tau \to e \gamma)$
is lost. That is Fig.~(\ref{alphaMMAA}-b) contains points with
${\rm Br}(\tau \to e \gamma)\sim 10^{-7}$, $-0.9<A_P<0.9$ and
$d_e< 10^{-29}~e~{\rm cm}$. The presence of these points can be
explained by the fact that when $A_{e\tau}$ and $(m_L^2)_{e \tau}
$ are nonzero but $A_{\tau e}=(m_R^2)_{e\tau}=0$ (or equivalently,
when $A_{\tau e}$ and $ (m_R^2)_{e\tau}\ne 0$ but
$A_{e\tau}=(m_L^2)_{e\tau}=0$) $d_e$, despite large
$\phi_{A_\tau}$, remains zero but $ A_L$ can be of order of $A_R$
which yields $-0.9<A_P<0.9$. As a result, without independent
knowledge of the ratios of LFV elements, we cannot derive any
conclusive bound on $\phi_{A_\tau}$ even if we find $-0.9<A_P<0.9$
and $10^{-8}<{\rm Br}(\tau \to e \gamma)$. We have repeated the
same analysis for other benchmarks and the results seem to be
robust against changing the mass spectrum.

As explained earlier, some models predict a certain pattern for
LFV. For example, within the framework of the seesaw mechanism
embedded in the constrained MSSM, we expect the LFV  to be induced
mainly on the left-handed sector \cite{borzumati}. That is we
expect $(m_R^2)_{e \tau}\ll (m_L^2)_{e \tau}$ and
$A_{e\tau}/A_{\tau e}\sim m_e/m_\tau\ll 1$. This model predicts
$A_P=-1$.  On the contrary, within the supersymmetric SU(5) GUT
model without right-handed neutrinos, the LFV is induced only on
the right-handed sector \cite{su(5)} which implies $A_P=1$. For
both of these cases, $d_e$ induced by $\phi_{A_\tau}$ is
negligible.

\begin{figure}
\begin{center}
\centerline{\epsfig{figure=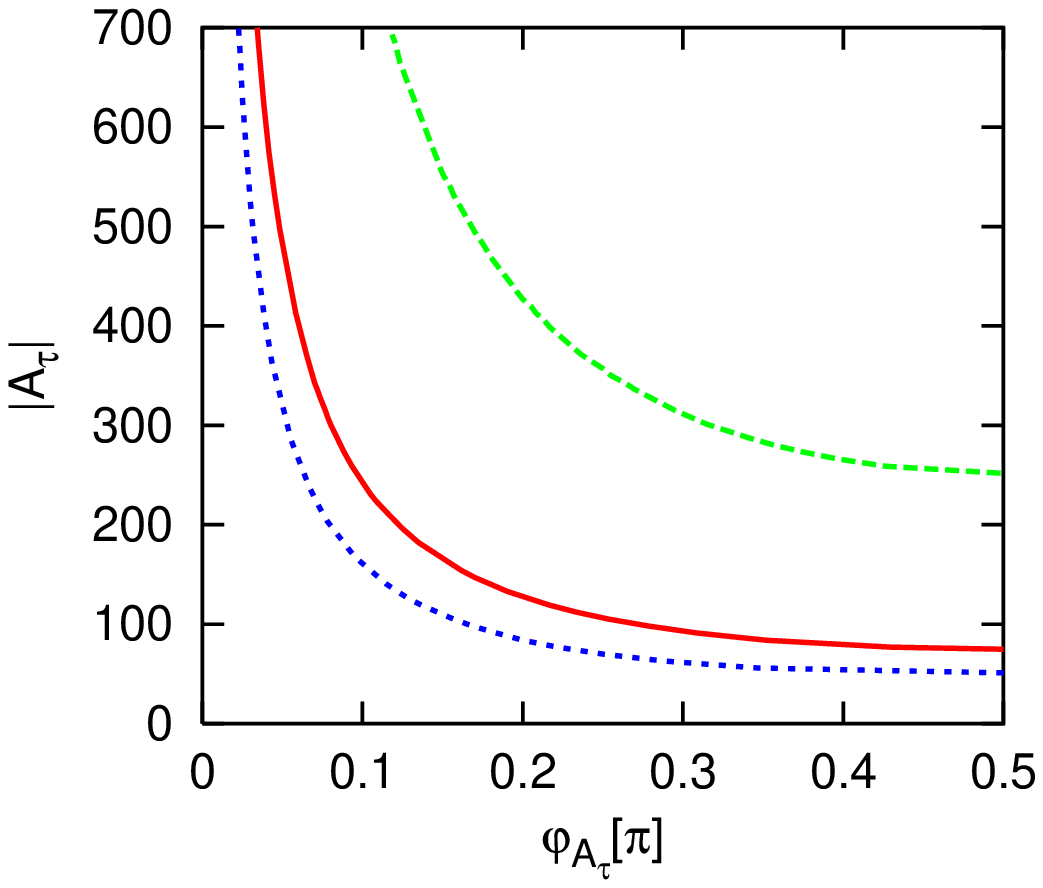,width=7.5cm}\hspace{5mm}\epsfig{figure=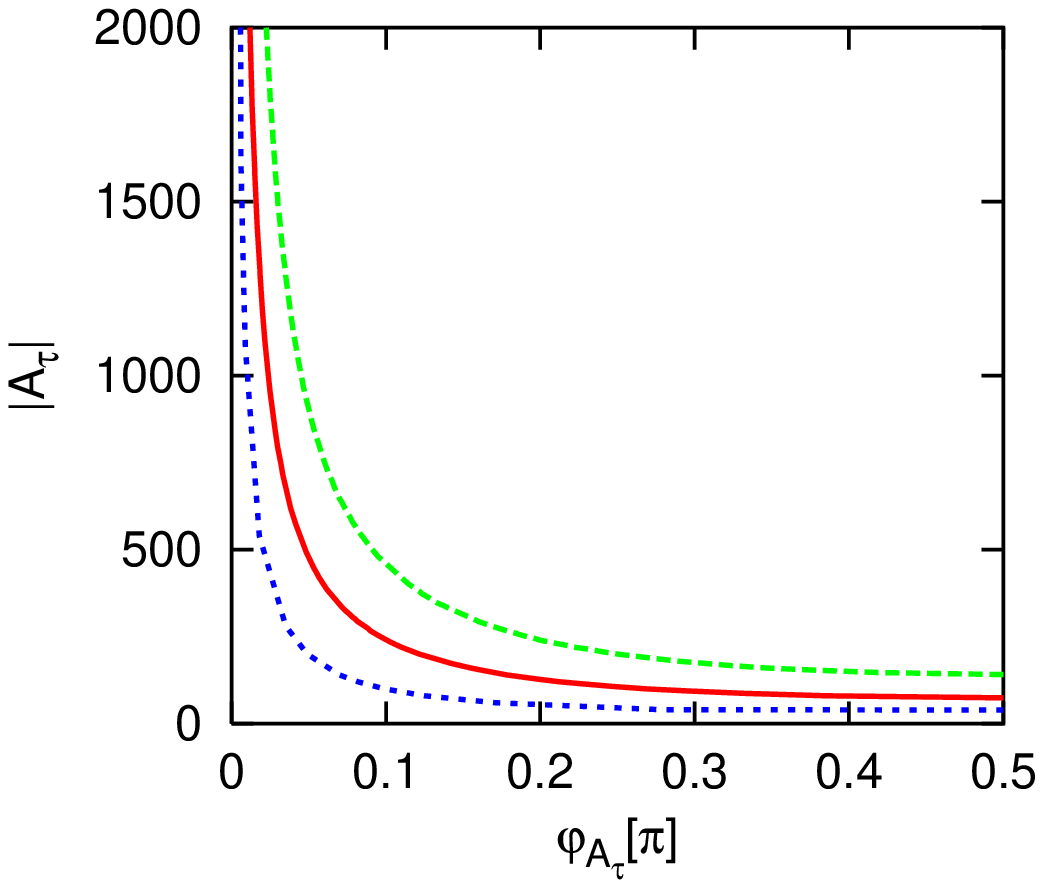,width=7.5cm}}
\centerline{\vspace{1.cm}\hspace{0.5cm}(a)\hspace{7cm}(b)}
\end{center}
\caption{Contour plots  for $d_e$=$1.4\times10^{-27}$. a) The
$\alpha$ benchmark proposed in \cite{NUHMbenchmark} is taken as
the input.  The dotted blue curve  corresponds to $(m_L^2)_{e
\tau}$=$3500~{\rm GeV}^2$, $(m_R^2)_{e \tau}$=$1500~{\rm GeV}^2$
and $(m^2_{LR})_{\tau e}$=$(m^2_{LR})_{e\tau}$=0 and the thick
solid red curve corresponds to $(m_L^2)_{e \tau}$=$3000~{\rm
GeV}^2$, $(m_R^2)_{e \tau}$=$1000~{\rm GeV}^2$,
$(m^2_{LR})_{e\tau}$=$300~{\rm GeV}^2$ and $(m^2_{LR})_{\tau
e}$=$100~{\rm GeV}^2$. The dashed green curve corresponds to
$(m_L^2)_{e \tau}$=$3000~{\rm GeV}^2$, $(m_R^2)_{e
\tau}$=$100~{\rm GeV}^2$, $(m^2_{LR})_{e\tau}$=$10~{\rm GeV}^2$
and $(m^2_{LR})_{\tau e}$=$500~{\rm GeV}^2$. b) The $\delta$
benchmark proposed in \cite{NUHMbenchmark} is taken as the input:
$|\mu|=920$~GeV, $m_0=500$~GeV, ${\rm M}_{1/2}=750$~GeV and $\tan
\beta=10$.  The dotted blue curve corresponds to $(m_L^2)_{e
\tau}$=$7\times 10^4~{\rm GeV}^2$ and $(m_R^2)_{e \tau}$=$2\times
10^4~{\rm GeV}^2$ and the thick red curve corresponds to
$(m_L^2)_{e \tau}$=$2\times 10^4~{\rm GeV}^2$, $(m_R^2)_{e
\tau}$=$3\times 10^4~{\rm GeV}^2$, $(m^2_{LR})_{e\tau}$=$3000~{\rm
GeV}^2$ and $(m^2_{LR})_{\tau e}$=$7000~{\rm GeV}^2$. The dashed
green curve corresponds to $(m_L^2)_{e \tau}$=$1\times 10^5~{\rm
GeV}^2$, $(m_R^2)_{e \tau}$=$3000~{\rm GeV}^2$,
$(m^2_{LR})_{e\tau}$=$30~{\rm GeV}^2$ and $(m^2_{LR})_{\tau
e}$=$8000~{\rm GeV}^2$. } \label{AphiA}
\end{figure}
In the above discussion, we have used the bound on $\phi_{A_\tau}$
and on ${\rm Im} (A_\tau)$ interchangeably. To clarify the
relation between these two, Fig.~(\ref{AphiA}) has been presented
which shows curves of $d_e=1.4\times 10^{-27}~e~{\rm cm}$ (the
present bound) for the $\alpha$ and $\delta$ benchmarks and
various values of the off-diagonal elements. The values of the LFV
elements are chosen in a range to obtain ${\rm Br}(\tau \to e
\gamma)$ close to the present bound; {\it i.e.,} $10^{-8}<{\rm
Br}(\tau \to e\gamma)<10^{-7}$. Each curve can be considered as
the upper bound on $\phi_{A_\tau}$. These figures also confirm
that when  there is a hierarchy between the left and right LFV
elements, the bounds are weaker. As expected, the curves have a
shape close ${\rm Im}(A_\tau)=|A_\tau| \sin \phi_{A_\tau}={\rm
cte} .$

 In summary, within
a model that $A_{ij}=0$, if ${\rm Br}(\tau \to e \gamma)$ turns
out to be close to its present bound and $A_P$ deviates from +1
and -1, the  bound on $d_e$ puts a strong bound on  Im$(A_\tau)$.
However, if $A_P=\pm 1$, the bound on $d_e$ can be explained by a
hierarchy between the $(m_{L}^2)_{\tau e}$ and $(m_{R}^2)_{\tau
e}$ elements instead of by the smallness of Im$(A_\tau)$. Similar
discussion holds for the scenario in which $(m_R^2)_{e
\tau}=(m_L^2)_{e\tau}=0$ and instead $A_{e\tau}$ and $A_{\tau e}$
are nonzero: while for $A_P\ne \pm 1$, the bound on $d_e$ can
severely restrict $\phi_{A_\tau}$, for $A_P=\pm 1$ we cannot
obtain any bound on $\phi_{A_\tau}$ from $d_e$. However, within a
scenario that $A_{e \tau}$, $A_{\tau e}$, $(m_{R}^2)_{\tau e}$ and
$(m_{L}^2)_{\tau e}$ are all large, we cannot derive any bound on
$\phi_{A_\tau}$ even if $A_P\ne \pm 1$. Thus, in order to derive a
conclusive bound on $\phi_{A_\tau}$, one has to resolve these
degeneracies seeking help from an experiment other than the rare
$\tau$ decay. Studying LFV signals at a $e^- e^+$ collider with
energy of center of mass of a few hundred GeV  can help in this
direction \cite{ilc}. In this paper we have concentrated on the
possibilities that the ongoing experiments can bring about.
Studying the possibilities with ILC is beyond the scope of the
present paper.

\section{Degeneracies between different sources of CP-violation}
In the previous section, we had assumed that the only source of
CP-violation is the phase of $A_\tau$. However, within the
framework of general MSSM, there are multiple sources  of
CP-violation.  In the basis described in Sec. 2, these phases
include the phases of $A_e$, the $\mu$-term and $M_1$ (the Bino
mass) that can contribute to $d_e$ regardless of the conservation
or violation of LF. Within the scenario considered in this paper,
in addition to these sources, the phases of $A_{e \tau}$, $A_{\tau
e }$, $(m_L^2)_{e \tau}$ and $(m_R^2)_{e \tau}$ can be also
considered as independent sources of CP-violation that can
contribute to $d_e$. If we assume that only one of these various
phases is nonzero, the present bound on $d_e$ can be interpreted
as a strong bound on the nonzero phase. However, in general when
more than one phase is present, the effects of different phases
can cancel each other \cite{cancelation}. Moreover, if the
forthcoming searches report a nonzero $d_e$, without additional
information, we cannot disentangle the source of CP-violation. In
this section, we discuss the degeneracies between different
sources of CP-violation with special emphasis on the possibility
of cancelation.

\begin{figure}
\psfig{figure=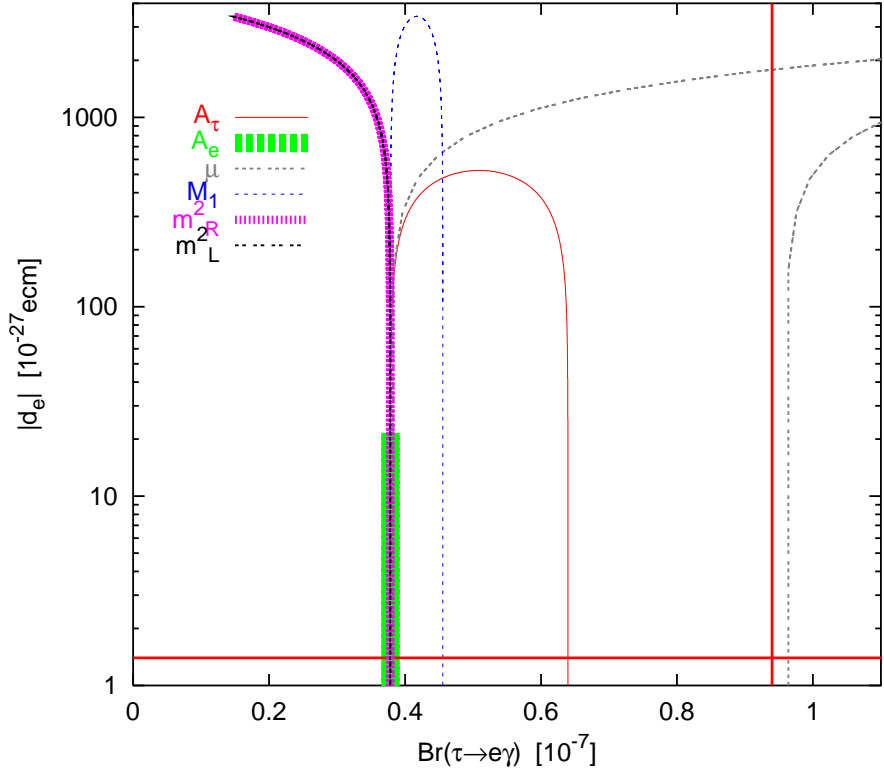,bb= 50 25 553 495, clip=true, height=4.5
in} \caption{$d_e$ versus ${\rm Br}(\tau\rightarrow e\gamma)$ as
the CP-violating phases vary between zero and $\pi$. The input
parameters correspond to the $\alpha$ benchmark proposed in
\cite{NUHMbenchmark}:$|\mu|=375$~GeV, $m_0=210$~GeV, ${\rm
M}_{1/2}=285$~GeV and $\tan \beta=10$ and we have set
$A_{\tau}$=$A_e$=500~GeV. All the LFV elements of the slepton mass
matrix are set zero except that $|(m_L^2)_{e \tau}|$=$3500~{\rm
GeV}^2$ and $|(m_R^2)_{e \tau}|$=$15000~{\rm GeV}^2$. To draw the
curves all phases are set zero except one that varies between 0
and $\pi$. As illustrated in the legend of the figure, the thin
solid red curve, dotted grey curve and light blue dashed  curve
respectively correspond to varying phases of $A_{\tau}$, $\mu$ and
$M_1$. The thin black and thick pink dotted curves correspond to
the phases of $(m_L^2)_{e \tau}$ and $(m_R^2)_{e \tau}$ which for
$A_{e\tau}=A_{\tau e}$ lie over each other. The thick green
vertical line stretching up to $d_e$=$2\times10^{-26}~e~{\rm cm}$
depicts the effect of  the phase of $A_e$. The horizontal line at
$1.4 \times 10^{-27}~e~{\rm cm}$ depicts the present experimental
limit \cite{pdg} and the vertical line shows the  present
experimental bound on ${\rm Br}(\tau\rightarrow e\gamma)$ at $9.4
\times 10^{-8}$ \cite{banerjee}.} \label{Pi0noLR}
\end{figure}

\begin{figure}
\psfig{figure=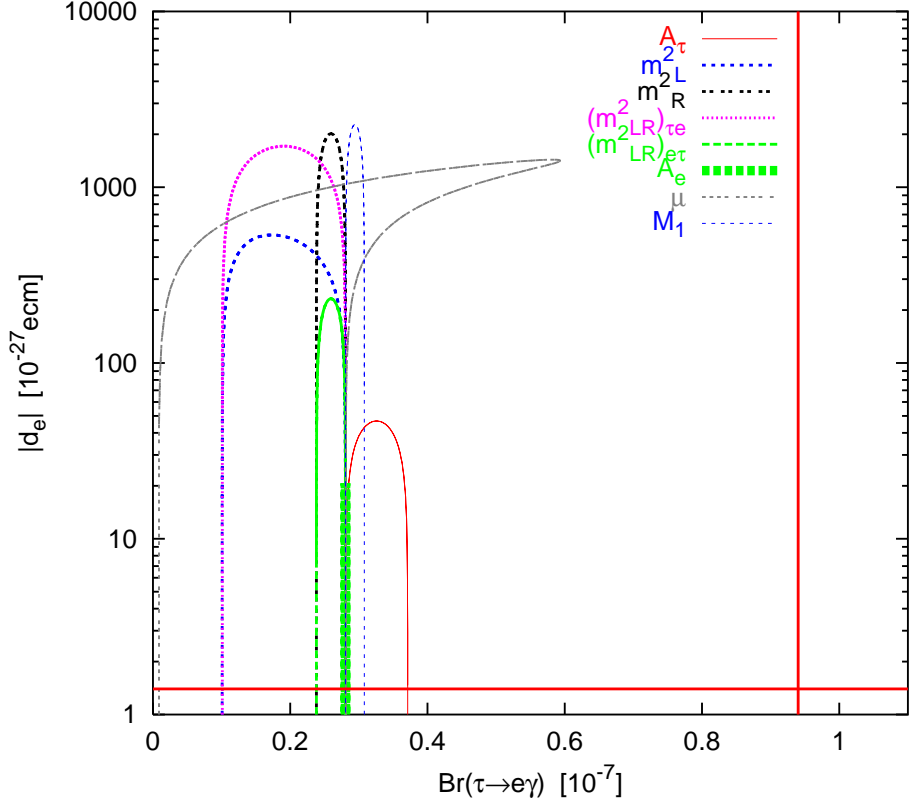,bb= 50 45 553 501, clip=true, height=4.5
in} \caption{Similarly to Fig.~\ref{Pi0noLR} except that here we
have set $|(m_L^2)_{e \tau}|=1000~{\rm GeV}^2$, $|(m_R^2)_{e
\tau}|$=$5000~{\rm GeV}^2$ and
$|(m^2_{LR})_{e\tau}|=|(m^2_{LR})_{\tau e}|=300~{\rm GeV}^2$. The
thin solid red curve, light dash-dotted grey curve, light blue
dashed  curve, solid dashed dark blue curve and thick black dotted
curve respectively correspond to the varying phase of $A_{\tau}$,
$\mu$, $M_1$, $(m_L^2)_{e \tau}$ and $(m_R^2)_{e \tau}$. The thick
green vertical line stretching up to
$d_e$=$2.1\times10^{-26}~e~{\rm cm}$ depicts the effect of the
phase of $A_e$. The light pink curve and thin green curve
respectively correspond to phases of $(m^2_{LR})_{\tau e}$ and
$(m^2_{LR})_{e\tau}$. The horizontal line at $1.4 \times
10^{-27}~e~{\rm cm}$ depicts the present experimental limit
\cite{pdg} and the vertical line  at $9.4 \times 10^{-8}$ shows
the present experimental bound on ${\rm Br}(\tau\rightarrow
e\gamma)$ \cite{banerjee}.} \label{Pi0LR}
\end{figure}

\begin{figure}
\psfig{figure=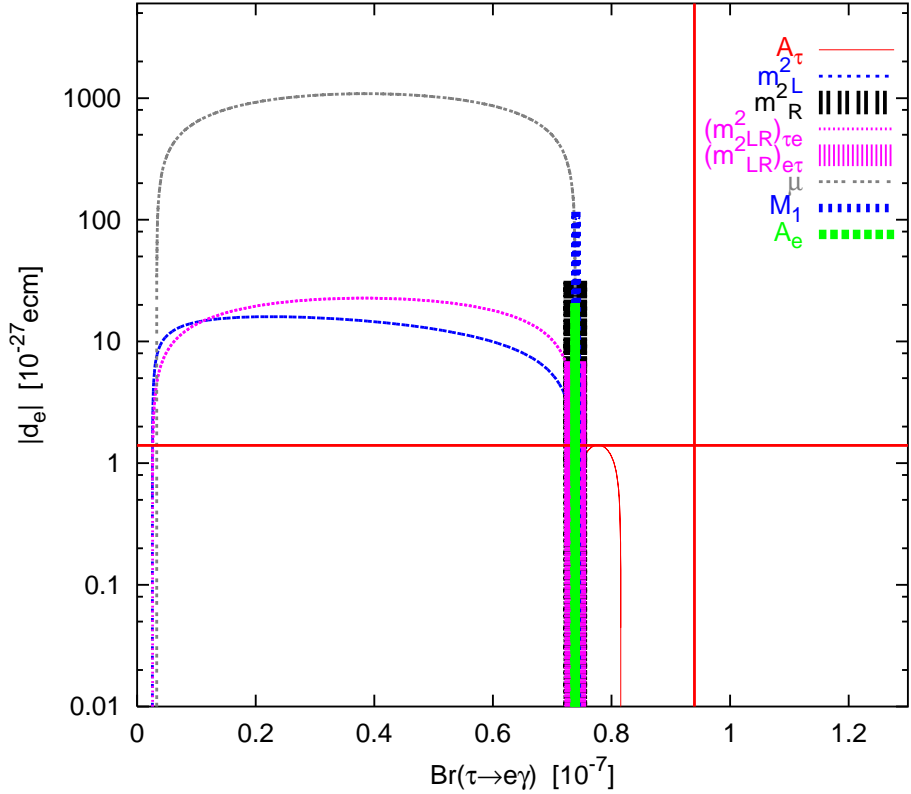,bb= 50 45 553 501, clip=true, height=4.5
in}\caption{Similarly to Fig.~\ref{Pi0noLR} except that here
$|(m_L^2)_{e \tau}|=3000~{\rm GeV}^2$, $|(m_R^2)_{e \tau}|=50~{\rm
GeV}^2$, $|(m^2_{LR})_{e\tau}|$$=|A_{e\tau}|\langle
H_d\rangle$=$3~{\rm GeV}^2$ and $|(m^2_{LR})_{\tau e}|$$=|A_{ \tau
e}|\langle H_d\rangle$=$400~{\rm GeV}^2$. The thin solid red
curve, light dotted grey curve, thin dotted blue curve and thin
dotted pink curve respectively correspond to varying phases of
$A_{\tau}$, $\mu$, $(m_L^2)_{e \tau}$ and $(m^2_{LR})_{\tau e}$.
The pink, green, black and dark blue thick vertical lines at ${\rm
Br} (\tau \to e\gamma)=7.5\times 10^{-8}$ (which reach up
$d_e=6.8\times 10^{-27},2\times 10^{-26},3.2\times
10^{-26},1.1\times 10^{-25}~e~{\rm cm}$) depict $d_e$ versus ${\rm
Br}(\tau \to e\gamma)$ as the phases of respectively
$(m_{LR}^2)_{e\tau}$, $A_e$, $(m_R^2)_{e\tau}$ and $M_1$ vary
between 0 and $\pi$. The horizontal line at $d_e=1.4\times
10^{-27}~e~{\rm cm}$ and the vertical line at ${\rm Br}(\tau \to e
\gamma)=9.4\times 10^{-8}$ show the present experimental
limits~\cite{pdg,banerjee}.} \label{Pi0hierarchy}
\end{figure}

 Figs.
(\ref{Pi0noLR}-\ref{Pi0hierarchy}) display the degeneracies
between possible CP-violating phases. To draw these figures, we
have inserted the mass spectrum of the $\alpha$ benchmark proposed
in \cite{NUHMbenchmark} (see caption of Fig. \ref{alphaMM} for the
values of the relevant parameters) and we have set
$A_\tau=A_e=500~{\rm GeV}$. Each of   Figs.
(\ref{Pi0noLR}-\ref{Pi0hierarchy}) corresponds to a different set
of absolute values for the  LFV elements $(m_L^2)_{e \tau}$,
$(m_R^2)_{e\tau}$, $A_{\tau e}$ and $A_{e\tau}$. Each curve in
these figures shows $d_e$ versus ${\rm Br}(\tau \to e \gamma)$ as
a certain CP-violating phase  varies from zero to $\pi$ while the
rest of the phases are set to zero. As expected all the curves
converge at $d_e=0$ which corresponds to the zero value of the
varying phase. As the value of the varying phase reaches $\pi/2$,
$d_e$ obtains its maximum value so the peak of each curve
corresponds to the varying phase equal to $\pi/2$. The horizontal
lines at $d_e=1.4 \times 10^{-27}~e~{\rm cm}$ in the figures show
the present upper bound on $d_e$ and the vertical lines at ${\rm
Br}(\tau \to e \gamma)=9.4\times 10^{-8}$ show the present bound
on ${\rm Br}(\tau \to e \gamma)$.
 In the following, we  discuss these figures one by
one.

Drawing  Fig.~\ref{Pi0noLR}, we have set
$|(m_L^2)_{e\tau}|=3500~{\rm GeV}^2$,
$|(m_R^2)_{e\tau}|=15000~{\rm GeV}^2$ and $A_{e\tau}=A_{\tau
e}=0$. The CP-violating phases that can contribute to $d_e$
include $\phi_{A_e}$, $\phi_{A_\tau}$, $\phi_\mu$, $\phi_{M_1}$
and the phases of $(m_L^2)_{e\tau}$ and $(m_R^2)_{e\tau}$. The
thick vertical line at ${\rm Br}(\tau \to e \gamma)=3.8\times
10^{-8}$ corresponds to the variation of $\phi_{A_e}$ in
$[0,\pi]$. This line shows that ${\rm Br}(\tau \to e \gamma)$ does
not significantly change as $\phi_{A_e}$ varies. The reason is
that the effect of $A_e$ on ${\rm Br}(\tau \to e \gamma)$ is much
smaller  than the dominant effect. The line associated with
$\phi_{A_e}$ (the thick line) reaches values of $d_e$ up to one
order of magnitude higher than the present bound on $d_e$ which
means if $\phi_{A_e}$ is the only contributor to $d_e$, it cannot
be larger than ${\cal O}(0.1)$. This bound is similar to the bound
in the LF conserving case. Notice that the effects of the rest of
phases can exceed the maximal contribution from $\phi_{A_e}$ by
more than one order of magnitude. In this figure, the curves
associated with the phases of $(m_L^2)_{e\tau}$ and
$(m_R^2)_{e\tau}$, which are depicted by black and pink dotted
curves, coincide. This observation is valid as long as
$A_{e\tau}=A_{\tau e}=0$ because the diagram shown in
Fig.~\ref{massinsertion} -- which in this case is the only diagram
contributing to $d_e$ -- is sensitive only to the relative phase
of $(m_L^2)_{e\tau}$ and $(m_R^2)_{e\tau}$. Another peculiar
feature of Fig.~(\ref{Pi0noLR}) is that the contribution of
$\phi_{M_1}$ to $d_e$ can exceed the maximum $d_e$ from
$\phi_\mu$. This is opposite to the LF conserving case in which
the effect of $\phi_\mu$ is larger because, while $\phi_{M_1}$ can
induce $d_e$ only through the subdominant neutralino-exchange
diagram, $\phi_\mu$ can induce EDM also through the dominant
chargino-exchange diagram. In contrast to the LF conserving case,
in the case of Fig.~(\ref{Pi0noLR}) the neutralino exchange
diagram dominates because as explained in the previous section,
once we turn on the $e\tau$ elements, the neutralino-exchange
diagram contributing to $d_e$ is enhanced by a factor of
$m_\tau/m_e$. As a result, the effect of $\phi_{M_1}$ is enhanced.

Now let us discuss the degeneracy and the possibility of
cancelation among different contributions. Replacing a phase with
its opposite value, its contribution to $d_e$ will change sign. As
a result if we find two phases whose  contributions  to $|d_e|$
have the same values, we can conclude that cancelation can take
place for at least one pair of values.
 Fig.~\ref{Pi0noLR} shows that the curve associated with
$\phi_{A_e}$ has a complete overlap with the low phase  part of
the other curves. That is for any value of $\phi_{A_e}$, there is
a value for other phases which can mimic the effect of
$\phi_{A_\tau}$. Thus, if the future EDM searches report a nonzero
value for $d_e$, there will be an ambiguity in interpretation of
the observation in terms of the phases. Cancelation is another
consequence of this overlap.   This figure shows that turning on
more than one nonzero phase, cancelation can make even the maximal
value of $\phi_{A_e}$ consistent with the present bound on $d_e$.
That is the contribution of $\phi_{A_e}=\pi/2$ can be canceled out
by the effect(s) of any  of the phases  $\phi_{M_1}$, $\phi_\mu$
or the phases of $(m_L^2)_{e\tau}$ and  $(m_R^2)_{e\tau}$ if these
phases are $\simeq \pi/500$.
 The contribution of
$\phi_{A_e}=\pm \pi/2$ can be also canceled out by the
contribution of $\phi_{A_\tau}$ if $|\phi_{A_\tau}|\simeq \pi/80$.

 Whereas the phase of $A_e$ can only
show up in $d_e$, the rest of phases can manifest themselves as
CP-odd effects at ILC \cite{ILC}. Moreover, $\phi_{M_1}$ and
$\phi_\mu$ can give  a detectable contribution to the EDM of the
neutron \cite{neutron}, mercury \cite{mercury} and deuteron
\cite{deuteron} through inducing chromoelectric dipole moments and
EDMs to the light quarks. Thus, from the experimental point of
view, cancelation between the effects of these phases is more
exciting as it can open up the possibility of large phases and
therefore CP-odd observable quantities  in  experiments other than
$d_e$ searches. Fig.~\ref{Pi0noLR} shows that  there are values of
$\phi_{M_1}$ and/or  the phases of $(m_{L}^2)_{e\tau}$ and/or
$(m_{R}^2)_{e\tau}$ whose contribution to $d_e$ can cancel even
the maximal effect from $\phi_\mu$. That is even $\phi_\mu=\pm
\pi/2$ is consistent with the bound on $d_e$ but in order to
cancel the effects down to the present bound a fine-tuning better
than 0.1\% is needed. In \cite{Bartl}, it was also shown that
turning on the LFV elements of mass matrix, the effect of
$\phi_\mu$ on $d_e$ can be canceled by the effects of the phases
of the off-diagonal elements. This result is obviously sensitive
to the largeness of the absolute values of the LFV elements which
are constrained by the null results of searches for the LFV rare
lepton decays. Bearing in mind that since \cite{Bartl}, these
bounds have significantly improved the above discussion can be
considered as an update and re-confirmation of the claim in
\cite{Bartl} in view of the recent bounds. Notice however that a
large $\phi_\mu$ can also give contributions to $d_n$ and $d_{\rm
Hg}$ exceeding their present bounds. The correlation between $d_e$
and $d_{{\rm Hg}}$ has been systematically studied  in
\cite{leptonquark}. To satisfy the bounds on $d_n$ and $d_{{\rm
Hg}}$ in the presence of a large $\phi_\mu$, there should be
another cancelation scenario at work in the quark sector. This
further suppresses the allowed parameter space; {\it i.e., a
double-folded} fine-tuning.

Let us now suppose that there is a symmetry or a mechanism that
sets $\phi_\mu=\phi_{M_1}=0$ so the bounds on $d_n$ and $d_{Hg}$
are naturally satisfied without the above mentioned {\it
double-folded} fine-tuning problem. Let us also suppose that
$\phi_{A_\tau}$ and the phases of $(m_{L}^2)_{e\tau}$ and
$(m_R^2)_{e\tau}$ are large. Fig.~\ref{Pi0noLR} shows that this
scenario is not ruled out by the $d_e$ bound because there is
still the possibility of cancelation between the contributions of
the nonzero phases. To cancel the effects of $\phi_{A_\tau}\simeq
\pi/2$ on $d_e$ down to the present bound, a fine-tuning better
than 1\% is required.

Fig.~(\ref{Pi0LR}) has an input similar to that of
Fig.~(\ref{Pi0noLR}) except that $A_{e \tau}$ and $A_{\tau e}$ are
set nonzero and smaller values for $(m_L^2)_{e\tau}$ and
$(m_R^2)_{e\tau}$ are chosen. Notice that unlike
Fig.~(\ref{Pi0noLR}) curves associated with the phases of
$(m_L^2)_{e\tau}$ and $(m_R^2)_{e\tau}$ split. The peak of the
$\phi_{A_\tau}$ curve in Fig.~(\ref{Pi0LR}) lies one order of
magnitude below that in Fig.~(\ref{Pi0noLR}). That is because the
absolute values of $(m_L^2)_{e\tau}$ and $(m_R^2)_{e \tau}$  in
Fig.~(\ref{Pi0LR}) are smaller. Had we set these elements larger,
the effect of $\phi_{A_\tau}$ on $d_e$  would have been larger but
also the value of  ${\rm Br}(\tau \to e \gamma) $ would have
increased. The rest of the argument for Fig.~(\ref{Pi0noLR}) holds
for Fig.~(\ref{Pi0LR}), too.

Fig.~(\ref{Pi0hierarchy}) displays the dependence of $d_e$ and
${\rm Br}(\tau \to e \gamma)$ on different phases for the case
that there is a hierarchy between the left and right LFV elements:
$|A_{e\tau}|\ll|A_{\tau e}|$ and $|(m_R^2))_{e\tau}|\ll
|(m_L^2)_{e\tau}|$. Because of this hierarchy, the effect of
$\phi_{A_\tau}$ on $d_e$ has dropped below the present bound which
is expected following the discussion in the previous section.
 The lines associated with the phases of $(m_R^2)_{e\tau}$  and
 $(m_{LR}^2)_{e\tau}$ appear as vertical lines which means ${\rm
 Br}(\tau \to e \gamma)$ does not depend on these phases. This is
 expected because $|(m_R^2)_{e\tau}|$ and  $|(m_{LR}^2)_{e\tau}|$
 are very small. However, the effects of their phases can still exceed
the present bounds. The figure also shows that ${\rm Br}(\tau \to
e\gamma)$ strongly depends on the phases of $(m_L^2)_{e\tau}$ and
$(m_{LR}^2)_{\tau e}$. The effect of $\phi_{A_e}$ is similar to
the previous cases. The  $\phi_{M_1}$ curve also appears as a
vertical line which means ${\rm Br}(\tau \to e \gamma)$  does not
strongly depend on $\phi_{M_1}$.  The effect of $\phi_{M_1}$ on
$d_e$ in case of Fig.~(\ref{Pi0hierarchy}) is one order of
magnitude smaller than the case of Fig.~(\ref{Pi0LR}) and, like
the LF conserving case, is smaller than the effect of $\phi_\mu$.
In contrast to Figs. (\ref{Pi0noLR}) and (\ref{Pi0LR}), in this
case $\phi_\mu\simeq \pi/2$ is ruled out by the bound on $d_e$
because the effects of other phases will not be large enough to
cancel the effect of $\phi_\mu\simeq \pi/2$. However, for
$\phi_\mu<\pi/30$ the effect of $\phi_{M_1}$ and for
$\phi_\mu<\pi/500$ the effects of the phases of LFV mass elements
as well as that $\phi_{A_e}$  can cancel the contribution from
$\phi_\mu$ to $d_e$.

\section{Concluding remarks}
In this paper, we have discussed the effects of the phase of
trilinear $A$-coupling of the staus, $\phi_{A_\tau}$, on $d_e$ in
the presence of nonzero LFV $e\tau$ elements of the slepton  mass
matrix. We have shown that for a large portion of the parameter
space consistent with the present bound on ${\rm Br}(\tau\to
e\gamma)$, the contribution of $\phi_{A_\tau}$ to $d_e$ can exceed
the present bound by several orders of magnitude. The effect of
$\phi_{A_\tau}$ on $d_e$ strongly depends on the ratios of the LFV
slepton masses $(m_L^2)_{e\tau}/(m_R^2)_{e\tau}$ and
$(m_{LR}^2)_{e\tau}/(m_{LR}^2)_{\tau e}$. In other words, for a
given  ${\rm Br}(\tau \to e\gamma)$ and $\phi_{A_\tau}=\pm \pi/2$,
$|d_e|$ can take any value between zero and a maximum which
depends on the value of ${\rm Br}(\tau \to e\gamma)$  [see
Figs.~(\ref{alphaMM}-a)-(\ref{alphaMMAA}-a)]. We have shown that
for specific case that $(m_{LR}^2)_{e \tau}=(m_{LR}^2)_{\tau e}=0$
[see Fig.~(\ref{alphaMM}-b)] or
$(m_{L}^2)_{e\tau}=(m_{R}^2)_{e\tau}=0$ [see
Fig.~(\ref{alphaAA}-b)], by measuring the asymmetry  $A_P$ defined
in Eq.~(\ref{apdef}) we can solve this ambiguity. However, in the
general case that all the $e\tau$ elements are nonzero, as shown
in Fig.~(\ref{alphaMMAA}), the correlation between $A_P$ and $d_e$
is lost and to solve the ambiguity, extra information is needed.

Assuming that $\phi_{A_\tau}$ is the only source of CP-violation
contributing to $d_e$ we have derived bounds on $\phi_{A_\tau}$
for various values of the LFV elements giving rise to ${\rm
Br}(\tau \to e\gamma)$ close to the present bound (see Figs.
\ref{forgotten} and \ref{AphiA}). We have then relaxed this
assumption and discussed the possibility of cancelation between
contributions of the different phases. We have shown that for
large $e \tau$ mass elements saturating the present bounds, the
effect of the phase of the Bino, $\phi_{M_1}$, on $d_e$ is
significantly enhanced which can be explained by the enhancement
of the effect of the neutralino exchange diagram by a factor of
$m_\tau/m_e$. Taking into account the new bounds on branching
ratios of the rare LFV tau decay, we have confirmed the results of
\cite{Reconciling} that with nonzero LFV effects cancelation
scenario makes large values of $\phi_{\mu}$ consistent with the
bound on $d_e$. We have discussed that the requirement to
simultaneously satisfy the bounds on $d_e$, $d_n$ and $d_{Hg}$ by
cancelation imposes a double-folded fine tuning problem.

We have shown that contributions from phases of $(m_L^2)_{e\tau}$,
$(m_R^2)_{e\tau}$, $(m_{LR}^2)_{e\tau}$ and $(m_{LR}^2)_{\tau e}$
can cancel the effect of $\phi_{A_\tau}$ on $d_e$. In summary,
although in case of large $e\tau$ elements saturating the bounds
from ${\rm Br}(\tau \to e \gamma)$, $\phi_{A_\tau}$ can induce a
large contribution to $d_e$, still the possibility of cancelation
and/or presence of a hierarchy between the LFV $e\tau$ mass matrix
elements make even a maximal $\phi_{A_\tau}$ consistent with the
$d_e$ bound  even if Br($\tau \to e \gamma$) is found to be close
to its present bound. Thus, still there is a hope to observe
CP-odd effects at ILC \cite{ILC}.

\section*{Acknowledgement} The authors would like
to thank M. M. Sheikh-Jabbari for careful reading of the
manuscript.

\section*{Appendix} In this appendix, we summarize the formulas
necessary for calculating $A_P$, ${\rm Br}(\tau \to e \gamma)$ and
$d_e$. In this paper, we are interested in large LFV $e\tau$
elements. In this parameter range, the mass insertion
approximation is not valid and one should work in the mass basis.
Here, we first derive the coupling of the sleptons to neutralinos
and charginos in the mass basis taking to account the CP-violating
phases and mixing. We then present the formulas for $A_L$ and
$A_R$ defined in Eq.~(\ref{effectiveL}) as well as for the formula
for $d_e$. Throughout this appendix we omit the spinorial indices
for simplicity.

 In the flavor basis, the mass terms of
$\widetilde{e}_L$  (the superpartners of the left-handed charged
leptons) and $\widetilde{e}_R$ (the superpartners of the
right-handed charged leptons) can be written as
\begin{equation}
L_{{\rm slepton}}=-\left(%
\begin{array}{cc}
  \widetilde{e}_L^\dagger& \widetilde{e}_R^\dagger \\
\end{array}%
\right)M_{\tilde{e}}^2\left(%
\begin{array}{c}
  \widetilde{e}_L \\
  \widetilde{e}_R \\
\end{array}%
\right)=
-\left(%
\begin{array}{cc}
  \widetilde{e}_L^\dagger& \widetilde{e}_R^\dagger \\
\end{array}%
\right)\left(%
\begin{array}{cc}
  m_L^2 & m_{LR}^{2\dag} \\
  m_{LR}^2 & m_R^2 \\
\end{array}\right)\left(%
\begin{array}{c}
  \widetilde{e}_L \\
  \widetilde{e}_R \\
\end{array}%
\right)
 \end{equation}
where $m_L^2$ and $m_R^2$ are $3\times3$ Hermitian matrices and
$m_{LR}^2$ is a  general complex $3\times3$ matrix. The elements
of these matrices are as follows:

\begin{equation} \label{mL}
(m_L^2)_{ij}=(m_{\widetilde{e}_L}^2)_{ij}+(m_{e}^2)_i\delta_{ij}+m_{Z}^2\cos2\beta(-\frac{1}{2}+
\sin^2\theta_W)\delta_{ij}
\end{equation}

\begin{equation} \label{mR}
(m_R^2)_{ij}=(m_{\widetilde{e}_R}^2)_{ij}+(m_{e}^2)_i\delta_{ij}-m_{Z}^2\cos2\beta\sin^2\theta_W
\delta_{ij}
\end{equation}
and
\begin{equation} \label{mLR}
(m_{LR}^2)_{ij}=m_{i}(A_{i}-\mu^*\tan\beta)\delta_{ij}+A_{ij}\langle
H_d \rangle
\end{equation}
%where $T_{3L(R)}^e$ and $ Q_{em}^e$ are weak isospin and electric
%charge, respectively.
where $m_{\widetilde{e}_R}^2$ and  $m_{\widetilde{e}_L}^2$   are
respectively  the right-handed and left-handed slepton soft
supersymmetry breaking  mass matrices at the electroweak  energy
scale and $A_{ij}$ is the trilinear $A$-coupling [see
Eq.~(\ref{MSSMsoft})].
% In
%general form, the above mass matrix is complex and off-diagonal
%which include mixing between different generation.
We can diagonalize the mass matrix of slepton by a $6\times6$
unitary matrix $U^l$ as
\begin{equation} [U^l
M_{\widetilde{e}}^2(U^l)^{-1}]_{xy}=m_{{\tilde{e}}_x}^2\delta_{xy}
\end{equation}
The slepton mass eigenstate in terms of the chiral weak eigenstate
are
\begin{equation}
\widetilde{e}_x=\sum_{i=1}^3[U_{x,i}^l\widetilde{e}_{Li}+U_{x,i+3}^l\widetilde{e}_{Ri}]
\end{equation}
Since in the MSSM no $\widetilde{\nu}_R$ exists, the neutrino mass
matrix will be a $3\times 3$ matrix whose elements can be written
as
\begin{equation}
(m_{\widetilde{\nu}}^2)_{ij}=(m_{\widetilde{e}_L}^2)_{ij}+(\frac{1}{2}m_{Z}^2\cos2\beta)\delta_{ij}
\end{equation}
The mass eigenstate, $\widetilde{\nu}_x$, is related to the weak
eigenstate, $\widetilde{\nu}_{Li}$, as
\begin{equation}
\widetilde{\nu}_{Li}=\sum^3_{x=1}U_{x,i}^{\nu*}\widetilde{\nu}_x
 \end{equation}
Let us now consider the  neutralino masses. The masses of
neutralinos in the weak basis can be written as
\begin{equation}
L_{{\rm neutralino}}=-\frac{1}{2}(\widetilde{X}^0)^T
M_{\widetilde{N}}\widetilde{X}^0+ {\rm H.c.},
 \end{equation}
 where $\widetilde{X}^0$=($\widetilde{B}$,
 $\widetilde{W}^0$, $\widetilde{H}_d^0$, $\widetilde{H}_u^0)$ and
\begin{equation}
M_{\widetilde{N}}= \left(
\begin{array}{cccc}
  M_1 & 0 & -m_Zc_\beta s_W & m_Zs_\beta s_W \\
  0 &  M_2 & m_Zc_\beta c_W & -m_Zs_\beta c_W \\
  -m_Zc_\beta s_W & m_Zc_\beta c_W & 0 & -\mu \\
  m_Zs_\beta s_W & -m_Zs_\beta c_W & -\mu & 0 \\
\end{array}%
\right).
\end{equation}
Here, $s_\beta=\sin\beta$, $c_\beta=\cos\beta$, $s_W=\sin\theta_W$
and $c_W=\cos\theta_W$.  The mass matrix $M_{\widetilde{N}}$ can
be diagonalized  as follows:
\begin{equation}
[{O_N}^* M_{\widetilde{N}}O_N^{-1}]_{AB}=M_{\tilde{\chi}_A^0}
\delta_{AB}
\end{equation}
where $O_N$ is a unitary matrix and $M_{\tilde{\chi}_A^0}$ are
real positive mass eigenvalues. The mass eigenstates,
$\widetilde{\chi}_{A}^0$, in terms of the weak eigenstates,
${\widetilde{X}_B}^0$, can be written as
\begin{equation}
\widetilde{\chi}_{A}^0=(O_N)_{AB}{\widetilde{X}_B}^0.
\end{equation}

 The chargino mass terms can be written as
\begin{equation}
L_{{\rm chargino}}=-\frac{1}{2}(\widetilde{X}^\pm)^T
M_{\widetilde{C}}\widetilde{X}^\pm+{\rm H.c.},
 \end{equation}
where $(\widetilde{X}^\pm)^T=(\widetilde{W}^+, \
{\widetilde{H}_u}^+, \ \widetilde{W}^-, \ \widetilde{H}_d^-)$ and
\begin{equation}
M_{\widetilde{C}}=
\left(%
\begin{array}{cc}
  0 & C^T \\
  C & 0 \\
\end{array}%
\right)
 \end{equation}
with
\begin{equation}
C=
\left(%
\begin{array}{cc}
  M_2 & \sqrt{2}s_\beta m_W \\
  \sqrt{2}c_\beta m_W & \mu \\
\end{array}%
\right).
 \end{equation}
The chargino mass matrix $C$ is a general complex matrix which can
be diagonalized as
\begin{equation}\label{cdiag}
U^c
CV^{c-1}=diag(|m_{\widetilde{\chi_1}^-}|,|m_{\widetilde{\chi_2}^-}|)
\end{equation}
where $U^c$ and $V^c$ are unitary matrices that satisfy the
following relations
\begin{equation}
V^c({C}^{\dag}C)V^{c-1}=diag(|m_{\widetilde{\chi_1}^-}|^2,|m_{\widetilde{\chi_2}}^-|^2)={U}^c(C{C}^{\dag}){U}^{c-1}.
\label{U}
\end{equation}
Notice that we have defined $U^c$ and $V^c$ in a way that the
elements of the diagonal matrix $U^cCV^{c-1}$ are real positive.
Eqs.~(\ref{cdiag},\ref{U}) are invariant under
\begin{equation}
U^c\rightarrow\left(%
\begin{array}{cc}
  e^{i\alpha_1} & 0 \\
  0 & e^{i\alpha_2} \\
\end{array}%
\right)U^c ,\ \ \ \ \ \ V^c\rightarrow\left(%
\begin{array}{cc}
  e^{i\alpha_1} & 0 \\
  0 & e^{i\alpha_2} \\
\end{array}%
\right)V^c.
\end{equation}
 Thus, there is an ambiguity in the definition of   $U^c$ and
$V^c$ but the final results do not depend on the unphysical phases
$\alpha_1$ and $\alpha_2$, as expected.

 The mass eigenstates  are related to the gauge eigenstates
through
\begin{equation}
\left(%
\begin{array}{c}
  \widetilde{\chi}_1^+ \\
  \widetilde{\chi}_2^+ \\
\end{array}%
\right)=V^c\left(%
\begin{array}{c}
  \widetilde{W}^+  \\
  \widetilde{H}_u^+  \\
\end{array}%
\right)\quad\left(%
\begin{array}{c}
  \widetilde{\chi}_1^- \\
  \widetilde{\chi}_2^- \\
\end{array}%
\right)=U^{c*}\left(%
\begin{array}{c}
  \widetilde{W}^-  \\
  \widetilde{H}_d^-  \\
\end{array}%
\right).
\end{equation}

Within the framework of the MSSM, the lepton-slepton-neutralino
coupling in the mass basis and in the Weyl representation can be
written as
\begin{equation}
L_{int}^{(n)}=\sum^6_{x=1}e_{Li}^\dag(N_{iAx}^R
)\widetilde{\chi}_A^{0\dag}\widetilde{e}_x+e_{Ri}^\dag(N_{iAx}^L
)\widetilde{\chi}_A^{0}\widetilde{e}_x+{\rm H.c.,}
 \end{equation}
where the couplings are
\begin{eqnarray}
N_{iAx}^R&=&-\frac{g_2}{\sqrt{2}}([-(O_N)_{A2}-(O_N)_{A1}\tan{\theta_W}]U_{x,i}^{l*}+\frac{m_{e_i}}{m_W\cos\beta}(O_N)_{A3}U_{x,i+3}^{l*})
\cr
N_{iAx}^L&=&-\frac{g_2}{\sqrt{2}}[2(O_N)_{A1}^*\tan{\theta_W}U_{x,i+3}^{l*}+\frac{m_{e_i}}{m_W\cos\beta}(O_N)_{A3}^*U_{x,i}^{l*}].
\end{eqnarray} The
lepton-slepton-chargino coupling can be  written as
\begin{equation}
L_{int}^{(c)}=\sum^3_{x=1}e_{Li}^\dag(C_{iAx}^{R}
)\widetilde{\chi}_A^{+\dag}\widetilde{\nu}_x+e_{Ri}^\dag(C_{iAx}^{L}
)\widetilde{\chi}_A^{-}\widetilde{\nu}_x
%+\nu_{i}^\dag(C_{iAx}^{\nu
%L})\widetilde{\chi}_A^{-\dag}\widetilde{e}_x+\nu_{i}^\dag(C_{iAx}^{\nu
%R} )\widetilde{\chi}_A^{-\dag}\widetilde{e}_x
{\rm +H.c.}
\end{equation}
where the couplings are
\begin{eqnarray}
C_{iAx}^{R}&=&-g_2U_{x,i}^{\nu*}V_{A,1}^{c} \cr
C_{iAx}^{L}&=&\frac{m_{e_{i}}}{\sqrt{2}m_W\cos\beta}g_2U_{x,i}^{\nu*}U_{A,2}^{c}
.
\end{eqnarray}

Let us now discuss the formulas for $(A_L)_{ij}$ and $(A_R)_{ij}$
defined as \be \label{kolli} e \epsilon_\alpha^\dagger m_\tau
\bar{e}_i\sigma^{\alpha \beta}q_\beta [(A_L)_{ij}
P_L+(A_R)_{ij}P_R]e_{j} +{\rm H.c.} \ee where $\sigma^{\alpha
\beta}=\frac{i}{2}[\gamma^\alpha,\gamma^\beta]$ and $q_\beta$ is
the four-momentum of the photon. $P_L$ and $P_R$ are respectively
the left and right projection matrices.
 Notice that $(A_L)_{e\tau}$ and $(A_R)_{e\tau}$ defined in (\ref{effectiveL}) are
 the $e \tau$  component of the $3\times 3$ matrices $(A_L)_{ij}$
  and $(A_R)_{ij}$. For the CP-conserving case the
formulas for $A_L$ and $A_R$ have been developed in \cite{hisano}.
We have rederived the formulae for the CP-violating case. It is
convenient to decompose $A_{L}$ and $A_R$ as follows
\begin{equation}
A_{L,R}=A_{L,R}^{(n)}+A_{L,R}^{(c)}
\end{equation}
where $A_{L,R}^{(n)}$ and $A_{L,R}^{(c)}$ respectively come from
neutralino-slepton and chargino-sneutrino loops. In terms of the
coupling in the mass basis we can write
\begin{equation}
\begin{split}
(A_L^{(n)})_{ij}=\sum_{A=1}^4 \sum_{x=1}^6
&\frac{1}{32\pi^2}\frac{1}{m_{\widetilde{e}_x}^2}[{N^{L}_{iAx}}{N^{L*}_{jAx}}\frac{1}{{6(1-y_{Ax})}^4}
\cr&\times(1-6y_{Ax}+3y^2_{Ax}+2y^3_{Ax}-6y^2_{Ax}\ln
y_{Ax})\cr&+{N^{L}_{iAx}}{N^{R*}_{jAx}}
\frac{M_{\widetilde{\chi}^0_A}}{m_{e_j}}\frac{1}{{(1-y_{Ax})}^3}
(1-y^2_{Ax}+2y_{Ax}\ln y_{Ax})],
\end{split}
\end{equation}
where $y_{Ax}=M^2_{\tilde{\chi}_A^0}/m^2_{\tilde{e}_x}$ and
\begin{equation}
\begin{split}
(A_L^{(c)})_{ij}=\sum_{A=1}^2\sum_{x=1}^3&-\frac{1}{32\pi^2}\frac{1}{m_{\widetilde{\nu}_x}^2}
[{C^{L}_{iAx}}{C^{L*}_{jAx}}\frac{1}{{6(1-z_{Ax})}^4}
\cr&\times(2+3z_{Ax}-6z^2_{Ax}+z^3_{Ax}+6z_{Ax}\ln
z_{Ax})\cr&+{C^{L}_{iAx}}{C^{R*}_{jAx}}\frac{M_{\widetilde{\chi}^-_A}}{m_{e_j}}\frac{1}
{{(1-z_{Ax})}^3} (-3+4z_{Ax}-z^2_{Ax}-2\ln z_{Ax})]
\end{split}
\end{equation}
$z_{Ax}={M^2_{\widetilde{\chi}^-_A}}/{m^2_{\widetilde{\nu}_x}}.$
Finally
\begin{equation}
A_{R}^{(n)}=A_{L}^{(n)}|_{L\leftrightarrow R} \ \ \
A_{R}^{(c)}=A_{L}^{(c )}|_{L\leftrightarrow R}.
\end{equation}
Notice that the forms of the above formulas are similar to those
in \cite{hisano}; however, the couplings $N^{R,L}$ and $C^{R,L}$
are slightly different because of the nonzero CP-violating phases.

Now, let us summarize the formula for $d_e$. It is also convenient
to decompose $d_e$ into the neutralino-exchange and
chargino-exchange contributions as follows:
$$d_e=d_e^{(n)}+d_e^{(c)}.$$ These contributions have been
extensively studied in the literature including in
\cite{edmformula} which give

$$d_e^{(c)}=-\frac{e}{(4\pi)^2}\sum_{A=1}^2\sum_{x=1}^3 {\rm
Im}(C^L_{eAx}C^{R*}_{eAx})
\frac{m_{\tilde{\chi}_A^-}}{m_{\tilde{\nu}_x}^2}A\left(\frac{m_{\tilde{\chi}_A^-}^2}{m_{\tilde{\nu}_x}^2}\right)$$

$$d_e^{(n)}=-\frac{e}{(4\pi)^2}\sum_{A=1}^4\sum_{x=1}^6 {\rm
Im}(N^L_{eAx}N^{R*}_{eAx}) \frac{m_{\tilde{\chi}_A^0}}
{m_{\tilde{e}_x}^2}B\left(\frac{m_{\tilde{\chi}_A^0}^2}{m_{\tilde{e}_x}^2}\right)$$
where
$$A(x)={1 \over 2 (1-x)^2} \left( 3-x+\frac{2 \ln
x}{1-x}\right)$$ and
$$B(x)={1 \over 2 (1-x)^2}\left( 1+x+\frac{2 x\ln
x}{1-x}\right).$$

\end{document}